\documentclass[preprint]{aastex}
\usepackage{float}
\usepackage{epsfig}
\usepackage{graphicx}
\usepackage{graphics}
\usepackage[latin1]{inputenc}
\usepackage{latexsym}
\newcommand{\nd}{\multicolumn{1}{c}{$\dots$}}

\shorttitle{SN~Ia in M101} \shortauthors{Zhang et al.}

\def\gsim{\;\lower4pt\hbox{${\buildrel\displaystyle >\over\sim}$}\;}
\def\lsim{\;\lower4pt\hbox{${\buildrel\displaystyle <\over\sim}$}\;}
\def\grls{\;\lower4pt\hbox{${\buildrel\displaystyle >\over <}$}\;}

\begin{document}

\title{Optical Observations of the Type Ia Supernova 2011fe \\ in M101 for Nearly 500 Days}

\author{Kaicheng Zhang\altaffilmark{1}, Xiaofeng Wang\altaffilmark{1}, JuJia Zhang\altaffilmark{2,3}, Tianmeng Zhang\altaffilmark{4}, Mohan Ganeshalingam\altaffilmark{5}, Weidong Li\altaffilmark{5,6}, Alexei V. Filippenko\altaffilmark{5}, Xulin Zhao\altaffilmark{1}, Weikang Zheng\altaffilmark{5}, Jinming Bai\altaffilmark{2,3}, Jia Chen\altaffilmark{1}, Juncheng Chen\altaffilmark{1}, Fang Huang\altaffilmark{1,7}, Jun Mo\altaffilmark{1}, Liming Rui\altaffilmark{1}, Hao Song\altaffilmark{1}, Hanna Sai\altaffilmark{1}, Wenxiong Li\altaffilmark{1}, Lifan Wang\altaffilmark{8}, and Chao Wu\altaffilmark{4}}

\altaffiltext{1}{Physics Department and Tsinghua Center for Astrophysics (THCA), Tsinghua University, Beijing, 100084, China;wang\_xf@mail.tsinghua.edu.cn}
\altaffiltext{2}{Yunnan Astronomical Observatory of China, Chinese Academy of Sciences, Kunming, 650011, China}
\altaffiltext{3}{Key Laboratory for the Structure and Evolution of Celestial Objects, Chinese Academy of Sciences, Kunming 650011, China}
\altaffiltext{4}{National Astronomical Observatory of China, Chinese Academy of Sciences, Beijing, 100012, China}
\altaffiltext{5}{Department of Astronomy, University of California, Berkeley, CA 94720-3411, USA}
\altaffiltext{6}{Deceased 12 December 2011}
\altaffiltext{7}{Astronomy Department, Beijing Normal University, Beijing, 100875, China}
\altaffiltext{8}{Physics and Astronomy Department, Texas A\&M University, College Station, TX 77843, USA}

\begin{abstract}

We present well-sampled optical observations of the bright Type Ia supernova (SN~Ia) SN 2011fe in M101. Our data, starting from $\sim16$ days before maximum light and extending to $\sim463$ days after maximum, provide an unprecedented time series of spectra and photometry for a normal SN~Ia. Fitting the early-time rising light curve, we find that the luminosity evolution of SN 2011fe follows a $t^n$ law, with the index $n$ being close to 2.0 in the $VRI$ bands but slightly larger in the $U$ and $B$ bands. Combining the published ultraviolet (UV) and near-infrared (NIR) photometry, we derive the contribution of UV/NIR emission relative to the optical. SN 2011fe is found to have stronger UV emission and reaches its UV peak a few days earlier than other SNe~Ia with similar $\Delta m_{15}(B)$, suggestive of less trapping of high-energy photons in the ejecta. Moreover, the $U$-band light curve shows a notably faster decline at late phases ($t\approx 100$--300 days), which also suggests that the ejecta may be relatively transparent to UV photons. These results favor the notion that SN 2011fe might have a progenitor system with relatively lower metallicity. On the other hand, the early-phase spectra exhibit prominent high-velocity features (HVFs) of O~I $\lambda$7773 and the Ca~II~NIR triplet, but only barely detectable in Si~II~6355. This difference can be caused either by an ionization/temperature effect or an abundance enhancement scenario for the formation of HVFs; it suggests that the photospheric temperature of SN 2011fe is intrinsically low, perhaps owing to incomplete burning during the explosion of the white dwarf.

\end{abstract}
\keywords{supernovae: general --- supernovae: individual (SN 2011fe)}

\section{Introduction}

SN 2011fe/PTF11kly was discovered in the nearby spiral galaxy M101 by the Palomar Transient Factory on August 24.167, 2011 (UT dates are used throughout this paper), and it was classified as a normal Type Ia supernova (SN~Ia; Nugent et al. 2011). SN 2011fe is one of the nearest Type Ia supernovae (SNe~Ia) discovered over the past three decades, with a Cepheid-based distance of about 6.4 $\pm$ 0.5 Mpc ($\mu =29.04\pm0.19$ mag; Shappee \& Stanek 2011). It is also one of the earliest detected SNe~Ia, within a few hours after the explosion (Nugent et al. 2011). Thus, the discovery of this object provides a rare opportunity to study the properties of a normal SN~Ia at both very early and very late phases.

There are two competing scenarios for the progenitors of SNe~Ia. One is a system consisting of a single white dwarf (WD) and a nondegenerate companion (Whelan \& Iben 1973); the other is a system consisting of two WDs (Iben \& Tutukov 1984). The fact that these explosions are linked to their birth environments suggests that SNe~Ia may arise from multiple classes of binary evolution (e.g., Wang et al. 2013). The early discovery of SN 2011fe leads to tight constraints on the nature of the progenitor of this particular SN~Ia.

The pre-explosion {\it Hubble Space Telescope (HST)} image of SN 2011fe ruled out luminous red giants and almost all helium stars as the mass-donating companion to the exploding WD (Li et al. 2011). Based on the early-time photometry of SN 2011fe, Bloom et al. (2012) set a limit on the initial radius of the primary star, $R_p \lesssim 0.02~{\rm R}_\odot$, as well as a limit on the size of the companion star, $R_c \lesssim 0.1~{\rm R}_\odot$. These studies suggest that for SN 2011fe, the companion star of the exploding WD is relatively compact, favoring a double-degenerate progenitor system. Using $Swift$ Ultraviolet/Optical Telescope (UVOT) observations, Brown et al. (2012) fit the $Swift$ UV light curves with the fireball model and concluded that there was no shock interaction with a nondegenerate companion. Similar conclusions were obtained from the analysis of radio and X-ray data (Chomiuk et al. 2012; Horesh et al. 2012). Based on deep Expanded Very Large Array (EVLA) radio observations, Chomiuk et al. (2012) constrained the density of the circumstellar material (CSM) and the mass-loss rate from the progenitor system, and they ruled out much of the parameter space associated with single-degenerate progenitor models for SN 2011fe. Horesh et al. (2012) used radio and X-ray observations of SN 2011fe to set a limit on the pre-explosion mass-loss rate of $\dot{M} \lesssim 10^{-8}(w/100~{\rm km~s}^{-1})$~M$_\odot$~yr$^{-1}$ for the progenitor system. They found that their data modestly disfavor the symbiotic progenitor model which involves a red-giant donor, but they cannot eliminate systems with an accreting main-sequence or subgiant star. Furthermore, the nondetection of hydrogen-rich material in the ejecta of SN 2011fe (i.e., with an upper limit of 0.001--0.003 M$_{\odot}$), inferred from its nebular spectra, further ruled out the possibility of having a hydrogen-rich star as the donor of the exploding WD (Shappee et al. 2013; Lundqvist et al. 2015; Graham et al. 2015a).

Besides studies of the progenitor itself, there are also many investigations of the observed properties of SN 2011fe. Richmond et al. (2012) present $BVRI$ photometry of SN 2011fe covering phases from $t = 2.9$ to 182 days after the explosion, finding a light-curve decline of $\Delta m_{15}(B) = 1.21\pm0.03$ mag. Munari et al. (2013) also published $B V R_C I_C$ light curves of SN 2011fe. Near-infrared (NIR) observations of SN 2011fe starting from 14 days before $B$-band maximum were obtained by Matheson et al. (2012), who also derived the Cepheid distance to M101 based on these data.

Studies of the earliest spectra were given by Parrent et al. (2012), Pereira et al. (2013), and Patat et al. (2013). Parrent et al. (2012) examined the high-velocity features (HVFs) and evolution of unburned material (carbon and oxygen) in the spectra. Pereira et al. (2013) presented spectrophotometric observations of SN 2011fe and explored the unburned carbon features in the spectra. Patat et al. (2013) studied the reddening along the line of sight toward SN 2011fe and concluded that this SN~Ia exploded in a ``clean'' environment. Multiple spectropolarimetric observations reveal that SN 2011fe has a low degree of continuum polarization, 0.2--0.4\% (Smith et al. 2011), indicating that the explosion was symmetric overall. The very late-time evolution of this object was recently reported by Taubenberger et al. (2015) and Graham et al. (2015a), based on optical spectra taken $\sim 1000$ days after the explosion.

In this paper, we present extensive photometry and spectroscopy covering phases from 16 days before to 463 days after $B$-band maximum light. Notwithstanding all the published data on SN 2011fe, our observations presented here are still a significant contribution to the literature and will aid future studies of SN 2011fe. With these data, we set better constraints on the explosion and progenitor properties of SN 2011fe. The observations are described in \S 2, light curves are presented in \S 3, and our spectra are shown in \S 4. A discussion is given in \S 5, and we conclude in \S 6.

\section{OBSERVATIONS}

\subsection{Optical Photometry}
The optical photometry presented here was obtained with the 0.8~m Tsinghua-NAOC Telescope (TNT; Wang et al. 2008; Huang et al. 2012), the 2.4~m Lijiang Telescope (LJT; Fan et al. 2015) of Yunnan Astronomical Observatory (YNAO), and the 0.76~m Katzman Automatic Imaging Telescope (KAIT; Filippenko et al. 2001). All of the data were reduced with standard IRAF routines. The instrumental magnitudes were converted to those of the Johnson $UBV$ (Johnson et al. 1966) and Kron-Cousins $RI$ (Cousins 1981) systems, based on transformation correlations established through observations performed on photometric nights. The $UBVRI$ magnitudes of 9 standard stars are listed in Table 1 (see Fig. 1 for the finder chart). The final flux-calibrated $UBVRI$ magnitudes of SN 2011fe are listed in Table 2.

\subsection{Optical Spectroscopy}

Our optical spectra of SN 2011fe were obtained by the 2.16~m telescope at Xinglong Observatory of NAOC and the 2.4~m LJT of YNAO; see Table 3 for the journal of observations. All spectra were reduced using standard IRAF routines and flux calibrated with spectrophotometric standard stars. The spectra were corrected for continuum atmospheric extinction at the two observatories, and telluric absorption lines were removed from the data.

\section{Light and Color Curves of SN 2011fe}

Figure 2 shows the $UBVRI$-band light curves of SN 2011fe from our observations; overplotted are the UV and NIR data (Brown et al. 2012; Matheson et al. 2012). The light curves resemble those of normal SNe~Ia, with a ``shoulder'' in the $R$ band and a prominent secondary maximum in the $I$ and NIR bands. For SN 2011fe, the peaks of the NIR and UV light curves appeared slightly earlier than in the $B$ band. Details of the light curves are described in the following subsections.

\subsection{The Light Curves}

A polynomial fit to the near-maximum light curves reveals that SN 2011fe reached a $B$-band maximum of $B_{\rm max} = 9.96\pm0.03$ mag on JD 2,455,814.98 $\pm$ 0.03 and a $V$-band maximum of $V_{\rm max} = 9.99\pm0.02$ mag on JD 2,455,816.92$\pm$0.03, very close to  $B_{\rm max} = 9.94 \pm 0.01$ mag on JD 2,455,815.01$\pm$0.06 and $V_{\rm max} = 9.98 \pm 0.02$ mag on JD 2,455,816.75 $\pm$ 0.06 as given by Pereira et al. (2013). We also derived the luminosity decline parameter $\Delta m_{15}(B) = 1.18\pm0.03$ mag and $B_{\rm max} - V_{\rm max} = -0.03 \pm 0.04$ mag, consistent with the values obtained by Richmond \& Smith (2012). From an empirical relation between intrinsic $B_{\rm max}-V_{\rm max}$ color and $\Delta m_{15}(B)$ (Phillips et al. 1999; Wang et al. 2009a), we can deduce $(B_{\rm max}-V_{\rm max})_0=-0.07 \pm 0.02$ mag for SN 2011fe. With the removal of the Galactic component, $E(B - V)_{\rm MW} = 0.008$ mag (Schlafly \& Finkbeiner 2011), we get $E(B - V)_{\rm host} = 0.032 \pm 0.045$ mag. Pereira et al. (2013) estimated the host-galaxy reddening as $E(B-V)_{\rm host} = 0.026 \pm 0.036$ mag from the photometric method and $E(B-V)_{\rm host} = 0.014 \pm 0.003$ mag from spectral data, while Tammann \& Reindl (2011) obtained $E(B-V)_{\rm host} = 0.030 \pm 0.060$ mag. Our result is consistent with these estimates within the quoted errors. Adopting a Cepheid distance modulus of $\mu = 29.04 \pm 0.19$ mag (6.4~Mpc; Shappee \& Stanek 2011) and correcting for the Galactic and host-galaxy extinction with $R_V = 3.1$, we derive absolute $B$ and $V$ magnitudes of $M_B = -19.24 \pm 0.19$ mag and $M_V = -19.17 \pm 0.19$ mag. Detailed photometric parameters of SN 2011fe are listed in Table 4.

Figure 3 shows comparisons of the near-maximum-light UV and optical light curves of SN 2011fe with those of well-observed normal SNe~Ia such as SN 2003du ($\Delta m_{15}(B) = 1.02$ mag; Stanishev et al. 2007), SN 2003hv ($\Delta m_{15}(B) = 1.61$ mag; Leloudas et al. 2009), SN 2005cf ($\Delta m_{15}(B) = 1.07$ mag; Wang et al. 2009a), SN 2011by ($\Delta m_{15}(B) = 1.16$ mag; Graham et al. 2015b; Song et al., in prep.), and SN 2012cg ($\Delta m_{15}(B) = 1.04$ mag; Munari et al. 2013; Marion et al. 2015). It is readily seen that the light curves of our comparison samples are similar around maximum light, except for SN 2003hv which exhibits a faster decline. Closer inspection reveals a faster rise in all bands for SN 2011fe compared with SN 2011by, although these two SNe~Ia have similar values of $\Delta m_{15}(B)$.

Figure 4 shows comparisons of the late-time light curves over the period from t $\approx$ +70 days to $\sim$ +500 days relative to $B$-band maximum. One can see that large differences emerge in the $U$ band; SN 2011fe declined more rapidly than
SN 2003du and even the fast decliner SN 2003hv. Past about 70 days after maximum light, the $U$-band decay rate of SN 2011fe is estimated to be $2.28\pm0.06$ mag (100 days)$^{-1}$, while the corresponding decline rates are $1.62\pm0.12$ mag (100 days)$^{-1}$ for SN 2003du (Stanishev et al. 2007) and $1.33\pm0.24$ mag (100 days)$^{-1}$ for SN 2003hv (Leloudas et al. 2009) at comparable phases. Although SN 2003du has small $\Delta m_{15}$(B), SN 2011fe still has a much faster decay rate compared to SN 2003hv.
The faster decay shown by the $U$-band light curve of SN 2011fe is thus likely caused by its ejecta having a relatively lower opacity in the UV (see also discussion in \S 5.3). Another possible factor is a difference in the magnetic field, which may lead to different amounts of positron trapping.

\subsection{The Color Curves}

Figure 5 shows the color curves of SN 2011fe, including the two UV minus $V$ colors ($uvw2-V$ and $uvw1-V$). Overplotted are the color curves of SNe 2003du, 2003hv, 2005cf, 2011by, and 2012cg. All of the color curves are corrected for reddening in both the Milky Way (Schlafly \& Finkbeiner 2011) and the host galaxies.

Inspection of the plot reveals that SN 2011fe has bluer $uvw1-V$ and $uvw2-V$ colors than the comparison SNe~Ia before maximum light (by $\sim0.2$ mag), consistent with its strong UV emission at early times. On the other hand, the $B - V$, $V - R$, and $V - I$ colors of SN 2011fe exhibit behaviors that are similar to those of the comparison SNe at early phases. The UV and optical color curves commonly showed an initial decline after the explosion and they reached their minimum values (i.e., bluest colors) around maximum light, followed by an increase toward redder colors until $t \approx +30$ days.

The subsequent evolution of the different color curves shows large scatter. The $uvw1 - V$ and $uvw2 - V$ colors become nearly constant during the period from $t \approx +30$ days to $t \approx +100$ days. Similarly, the $U-B$ color also showed a plateau feature during this phase, which then became redder in a linear fashion thereafter. At later phases, SN 2011fe is found to be progressively redder than SN 2003du and SN 2003hv in $U-B$, suggesting that its photospheric temperature drops more rapidly than that of SN 2003du and SN 2003hv. Meanwhile, $B - V$, $V - R$, and $V - I$ evolved toward bluer colors (declined by $\sim1.2$ mag) during the period from $t \approx +30$ days to $t \approx +200$ days. At $t \gtrsim +200$ days, the $B - V$ and $V - R$ colors showed a flat evolution while the $V - I$ color gradually became red again.

\subsection{The SED and Bolometric Light Curve}
Since we have photometry of SN 2011fe in the UV, optical, and NIR bands covering wavelengths of 1600--24,000~\AA, we can study its spectral energy distribution (SED) evolution by means of photometry. A rough SED can be constructed from the observed fluxes in various passbands at the same or similar epochs. The missing data can be obtained through interpolations of the neighboring datapoints whenever necessary. The observed fluxes are corrected for the reddening of Milky Way and the host galaxies.

Figure 6 shows the SEDs obtained at $t = -14$, $-7$, +1, +7, +16, and +31 days with respect to $B$-band maximum. One can see that the SEDs show a prominent deficit in the $uvm2$ band at all of these epochs, explained as line blending caused by iron-peak elements (e.g., Wang et al. 2009a). The SED of SN 2011fe is found to peak in the $B$ band at $t = -14$ days, and it then peaked in the $U$ band at $t =-7$ days and $t = +1$ days. After maximum light, the SED peak shifted quickly toward longer wavelengths. Around one month after maximum light, the emission in the $H$ band became stronger than that in the $J$ band, which is likely caused by the recombination of Fe~III. Compared to SN 2005cf (see Fig. 12 in Wang et al. 2009a), SN 2011fe seems to show a faster decrease in the photospheric temperature.

We constructed the bolometric light curve of SN 2011fe, as listed in Table 5, and compare it with that of SN 2003du, SN 2005cf, SN 2011by, and SN 2012cg, as shown in Figure 7. Owing to the lack of NIR data for SN 2011by, we assume it has the same NIR/optical ratio ($F_{\rm NIR}/F_{\rm optical})$ as SN 2005cf. With a peak luminosity of ($1.13\pm0.07)\times 10^{43}$ erg s$^{-1}$, we can deduce that the synthesized nickel mass is $M_{\rm Ni} = 0.57$~M$_{\odot}$ for SN 2011fe according to the Arnett law (Arnett 1982; Stritzinger \& Leibundgut 2005). In Figure 8, we plot the ratio of the UV (1600--3200~\AA) and NIR (9000--24,000~\AA) fluxes to the optical (3200--9000~\AA) for SN 2011fe and the comparison sample SN 2005cf, SN 2011by, and SN 2012cg. It can be seen that the UV/optical ratio ($F_{\rm UV}/F_{\rm optical})$ of SN 2011fe is comparable to that of SN 2005cf but apparently higher than that of SN 2011by and SN 2012cg, while the NIR/optical ratio ($F_{\rm NIR}/F_{\rm optical})$ is similar for these three SNe~Ia. We noticed that the $F_{\rm UV}/F_{\rm optical}$ ratio of SN 2011fe reached its peak one week earlier than that of SN 2005cf and also slightly earlier than that of SN 2012cg, suggestive of a shorter diffusion time for its higher-energy photons. Given a similar ejecta mass and expansion velocity, this difference implies that the ejecta of SN 2011fe have a lower opacity at shorter wavelengths.

\section{OPTICAL SPECTRA}

We have in total 35 optical spectra of SN 2011fe, obtained with the 2.4~m LJT of YNAO and the 2.16~m telescope of NAOC, spanning from $t = -16$ days to $t = +463$ days with respect to $B$-band maximum light. Figure 9 shows the complete spectral evolution. Detailed comparisons with some well-observed SNe~Ia at different epoches are shown in Figures 10 and 11.

\subsection{Temporal Evolution of the Spectra}
In Figure 10, we compare the spectra of SN 2011fe with those of SN 2003du (Stanishev et al. 2007), SN 2005cf (Wang et al. 2009a), SN 2011by (Graham et al. 2015b; Song et al., in prep.), and SN 2012cg (our own unpublished database) at four different epochs ($t \approx -14$ d, $-7$ d, 0 d, and +90 d with respect to $B$ maximum). At these phases (except for $t \approx 3$ months), spectra of SN 2011fe and the comparison sample show large differences in line profiles of some species, in particular the Si~II and Ca~II absorption.

Figure 10a shows the comparison of the spectra at $t \approx -14$ d. To identify the absorption features in the early-time spectra of SN 2011fe, we use SYNAPPS (Thomas et al. 2011) to fit two early-time spectra obtained at $t =-16$ d and $t =-12$ d, and the results are shown in Figure 12. One can see that the spectral features at these phases are dominated by intermediate-mass elements (IMEs) like calcium, silicon, oxygen, and magnesium, while the Fe~II/Fe~III absorptions are responsible for the troughs near 4300~\AA\ and 4800~\AA. Note that the fit to the $t = -12$ d spectrum looks better if Co~II and Ni~II are included. The notches near 6300~\AA\ and 7000~\AA\ can be attributed to C~II $\lambda$6580 and C~II $\lambda$7234 absorptions, respectively, which have also been identified by Parrent et al. (2012) and Pereira et al. (2013). Weak absorption from C~II is also detected in SN 2003du, SN 2005cf, SN 2011by, and SN 2012cg at earlier phases. For the Ca~II~NIR triplet lines, we found that high-velocity components are needed in order to get a better fit for SN 2011fe. At $t \approx -14$ d, the HVF of Si~II $\lambda$6355 is found to be very prominent in SN 2005cf and SN 2012cg, while it is weak in SN 2003du and does not seem to exist in SN 2011by and SN 2011fe. On the other hand, SN 2011fe displays two noticeable absorption features at $\sim 7300$~\AA\ and $\sim 7400$~\AA, and they can be identified as HVF and photospheric components of O~I $\lambda$7773 at velocities of $\sim 1.8 \times 10^{4}$ km~s$^{-1}$ and $\sim 1.3 \times 10^{4}$ km~s$^{-1}$, respectively (see Figure 15). Note that such O-HVFs are very weak or undetectable in the earliest spectra of our comparison SNe~Ia.

Figure 10b shows the comparison at $t \approx -7$~d. The most noticeable change in the spectra is the rapid evolution of the HVFs of O, Si, and Ca. For SN 2011fe and SN 2011by, the Ca-HVFs became almost invisible in the spectrum at this time. In contrast, the Ca-HVFs are still very strong in SN 2005cf and SN 2012cg, and the Si-HVF becomes very weak but still detectable in these SNe~Ia. On the other hand, at this phase, the HVF of O~I $\lambda$7773 still appears to be detectable in SN 2011fe. This indicates that SN 2011fe (and perhaps SN 2011by) has overall weaker Si-HVFs and Ca-HVFs compared to SN 2005cf and SN 2012cg, while it shows prominent O-HVFs.

In Figure 10c, we compare the near-maximum-light spectra. At $t \approx 0$~d, the spectrum of SN 2011fe has shown some evolution relative to the features exhibited at the earlier epochs, and the photospheric components of the O~I $\lambda$7773 and Ca~II~NIR triplet absorption are found to be stronger than in the comparison SNe~Ia. For SN 2005cf and SN 2012cg, the HVFs of the Ca~II~NIR triplet are still dominant over the photospheric components around maximum light.

In Figure 10d, we compare the spectra at $t \approx 3$ months. With the photosphere receding into the inner region, SN 2011fe and the comparison SNe~Ia exhibit quite similar spectral features. The absorption trough from the Ca~II~NIR triplet is still the dominant feature, and other main lines include Na~I, Fe~II, and Fe~III lines which develop into a highly similar profile in the nebular phase.

Two late-time nebular spectra, obtained with the YFOSC on day +233 and on day +463, are shown in Figure 11. Overplotted are the late-time spectra of SN 2003du, SN 2005cf, and SN 2012cg. One very late-time spectrum of SN 2011fe, taken on day +1034 (Taubenberger et al. 2015), is also shown for comparison. At such late phases, the spectra are dominated by forbidden lines of singly and doubly ionized iron-group elements, such as the [Fe~II] features at $\sim4400$~\AA, $\sim5200$~\AA, and $\sim7200$~\AA, [Fe~III] at $\sim4700$~\AA, and [Co~II] at $\sim4900$~\AA. These features are commonly seen in the comparison SNe~Ia at similar phases. Combining the t$\approx$ +233 d, +463 d, and +1034 d spectra, we notice that the [Fe~III] feature at $\sim 4700$~\AA, [Co~II] at $\sim 6000$~\AA, and [Fe~II]/[Ni~II] at $\sim7200$~\AA\ tend to become relatively weak with time, while the [Fe~II] features at $\sim4400$~\AA\ and $\sim5200$~\AA\ seem to show the opposite tendency.

\subsection{Evolution of Photospheric- and High-Velocity Features}

The well-sampled spectra of SN 2011fe can also allow us to study the evolution of the photospheric-velocity features (PVFs) and HVFs in the earliest spectra. From the evolution of these line profiles as displayed in Figure 13, one can hardly see the presence of an HVF in Si~II $\lambda$6355, while the HVFs are obviously seen in the Ca~II~NIR triplet. Note that our spectral sequence indicates that the O~I $\lambda$7773 line may have multiple absorptions formed at different velocities, as discussed below. For a thorough analysis of the PVFs and HVFs of Si and Ca in SNe~Ia, see Silverman et al. (2015; hereafter S15) and Zhao et al. (2015; hereafter Z15). In this subsection, we analyze the evolution of velocity and pseudo-equivalent width (pEW) of the absorptions of Si~II $\lambda$6355 and the Ca~II~NIR triplet for SN 2011fe, and compare the results with those from SN 2003du, SN 2005cf, 2011by, and SN 2012cg, as shown in Figures 14 and 15.

The velocity measured from the Si~II $\lambda$6355 absorption in near-maximum-light spectra is $1.04 \times 10^4$ km~s$^{-1}$ for SN 2011fe, showing that it belongs to the normal-velocity (NV) subclass of SNe~Ia in the classification scheme of Wang et al. (2009b). The velocity gradient, measured during the period from $t = 0$ d to $t = +10$ d, is found to be $\dot{v}=52.4$ km~s$^{-1}$ day$^{-1}$, suggesting that SN 2011fe can be put into the low-velocity gradient (LVG) subtype in the classification scheme of Benetti et al. (2005).

As discussed by S15 and Z15, the HVFs are usually prominent in the early-phase spectra. To detect the HVFs of SN 2011fe, we apply a two-component Gaussian function to fit the absorptions from Si~II $\lambda$6355 and the Ca~II~NIR triplet in the $t = -16$ d spectrum. We did not detect any significant HVF in the Si~II $\lambda$6355 absorption, but there are noticeably strong HVFs in the Ca~II~NIR triplet. The velocity evolution of the Si~II and Ca~II absorptions (both PVF and HVF) is shown in Figure 14, where we can see that the Ca-HVFs of SN 2011fe maintain a velocity of $\sim 2.0 \times 10^{4}$ km~s$^{-1}$ during the period from $t = -10$ d to $t = -5$ d with respect to $B$ maximum. This velocity plateau can be also seen in the Ca-HVFs of SN 2011by and SN 2012cg.

Figure 15 shows the pEW of Si~II $\lambda$6355 and the Ca~II~NIR triplet absorptions. It can be seen that the pEW of the Si-PVF of SN 2011fe reached a minimum at $\sim -6$ days, with a pEW of $\sim 80$~\AA. A similar trend can be seen in the comparison SNe, of which SN 2012cg has an overall weaker absorption. Our fit to the early-time Si~II absorption suggests that the Si-HVF is very weak or does not exist in SN 2011fe, consistent with previous analyses by S15 and Z15. In contrast, the HVFs are clearly detected in the Ca~II~NIR triplet of SN 2011fe and the four comparison SNe~Ia. This is consistent with the statistical result that the Ca-HVFs are more commonly seen in SNe~Ia than the Si-HVF. The absorption strength of Ca-HVFs decayed very quickly in SN 2011fe, changing from $\sim 250$~\AA\ at $t \approx -16$ days to $\sim 30$~\AA\ at $t \approx -5$ days. This is similarly seen in SN 2005cf, SN 2011by, and SN 2012cg.

In addition to the HVFs of the Ca~II~NIR triplet, the HVF of O~I~$\lambda$7773 can be detected in the earliest spectra of SN 2011fe, as shown in Figure 13. At $t \approx -16$ days, the HVF of O~I $\lambda$7773 is measured to have a velocity of $\sim 1.8 \times 10^{4}$ km~s$^{-1}$ and the photospheric component has a velocity of $\sim 1.4 \times 10^{4}$~km~s$^{-1}$. The photospheric velocity is consistent with that from Si~II and Ca~II lines. It is interesting to note that a second HVF of O~I~$\lambda$7773 may appear in the earlier spectra of SN 2011fe (see minor absorption marked by the dotted line), which has a velocity of $\sim 2.2 \times 10^{4}$~km~s$^{-1}$ (Zhao et al. 2016). The presence of oxygen at such a high velocity may naturally explain the formation of Ca-HVFs seen in SN 2011fe. Nugent et al. (2011) noticed that the O-HVF shows a rapid velocity decline from about 18,000 km s$^{-1}$ to 14,000 km s$^{-1}$ in the first two spectra (at $t = 1.2$~d and $t = 1.5$~d after explosion), and they attributed it to geometrical dilution during the early phases. However, it is more likely that the velocity variation they measured is actually related to different components of O~I~$\lambda$7773 absorption. From the SYNAPPS fit as shown in Figure 12, the feature with a velocity of $\sim 0.9 \times 10^{4}$ to $0.5 \times 10^{4}$~km~s$^{-1}$ may due to the blending of Mg~II and Si~II. The feature with a velocity of $\sim 3.3 \times 10^{4}$~km~s$^{-1}$ remains unknown to us.

\section{DISCUSSION}

\subsection{The Rising Light Curves}

The rising light curves are important, as they can determine the explosion time (Nugent et al. 2011), constrain the radius of the exploding star itself (Piro 2010; Bloom et al. 2012; Piro \& Nakar 2013), and test the scenario of luminosity evolution of SNe~Ia as well as other properties of their progenitor systems, such as interaction with a companion star (Kasen 2010). Note that there is a difference between the explosion time and the time when the SN begins to brighten (time of first light, $t_{\rm first}$) because of a possible dark phase immediately after the explosion (Piro \& Nakar 2013; Piro \& Nakar 2014); thus, we use $t_{\rm first}$ relative to $B$-band maximum in the following discussion. In the ``expanding fireball model,'' the early-time flux is thought to be $f \propto (t-t_{\rm first})^2$ (Riess et al. 1999; Conley et al. 2006). Assuming a more general form of the fireball model such as $f\propto (t - t_{\rm first})^{n}$, a recent study using 18 SNe~Ia yields a mean (but without stretch correction) rise time of $18.98 \pm 0.54$ d and a mean index of $n = 2.44 \pm 0.13$ (Firth et al. 2015). Some modified models, such as a broken power law, are also proposed for the luminosity evolution of SNe~Ia (e.g., Zheng et al. 2013).

For SN 2011fe, Nugent et al. (2011) obtained a $t_{\rm first}$ of MJD $55796.696\pm0.003$ ($t^2$) and $55796.687\pm0.014$ $(t^{2.01})$. Brown et al. (2012) estimated $t_{\rm first}$ ranging from MJD 55796.62 to MJD 55797.07 in different bands, using the earliest data from $Swift$ UVOT observations. With a bolometric light curve, Pereira et al. (2013) derived $t_{\rm first}$ as $55796.81\pm0.13$ for $t^2$ evolution and $t_{\rm first} = 55796.47\pm0.83$ for $t^n$, and $n = 2.21 \pm 0.51$.

Our extensive photometric data for SN 2011fe, starting within 1 day after the explosion, also enable better constraints on its luminosity evolution and first-light time. Using the observed data at $t < -10$ days and assuming a model of $f\propto (t - t_{\rm first})^{n}$ for the rising light curves, we find that the rising rate of the emission flux differs from $t^2$ evolution in all bands except $V$ and $I$, where the flux rises in a manner close to $t^2$ evolution (i.e., n = 2.27 and 2.33). Residuals of the fits are shown in the bottom panels of Figure 16, and the best-fit results are listed in Table 6. From the $t^{2}$ fit, we found that the first-light time ranges from $-17.37$ days ($U$ band) to $-17.82$ days ($V$ band), while the corresponding time estimated with the $t^n$ model spans from $-18.12$ days ($I$ band) to $-19.37$ days ($U$ band). In light of the goodness of the fit, one may conclude that the rising light curve of SN 2011fe differs from the $t^2$ evolution, especially in the $U$ and $B$ bands where a faster rise (and hence a larger index, $n > 2.7 $) is needed for a better fit, or the first-light times obtained in the $UB$ bands are smaller than those in the $VRI$ bands.

In principle, the first-light time (or explosion time) derived from different bands should have the same value, but this is not the case either for the $t^{2}$ fit or for the $t^{n}$ fit to the data. We thus refit the multi-band light curves of SN 2011fe simultaneously by forcing all the light curves to have the same $t_{\rm first}$. For the $t^2$ model, the combined fit gives $t_{\rm first} = -17.59 \pm 0.01$ (MJD 55796.89 $\pm$ 0.01) by using the data at $t < -10$ d, while for the $t^{n}$ model, the combined fit gives $t_{\rm first} = -18.18 \pm 0.29$ (MJD 55796.30 $\pm$ 0.29), with $n$ ranging from 2.25 (in the $V$ band) to 2.63 (in the $U$ band). We noticed that the $\chi^2$ for the $U$-band data is much larger than that for other bands, so we refit the data by restricting only to the $BVRI$ bands. We obtained $t_{\rm first} = -17.64 \pm 0.01$ (MJD 55796.84 $\pm$ 0.01) for the $t^2$ model, and $t_{\rm first} = -18.00 \pm 0.16$ (MJD 55796.48 $\pm$ 0.16) for the $t^{n}$ model, with $n$ ranging from 2.14 (in the $V$ band) to 2.43 (in the $B$ band). The detailed results are shown in the right panels of Figure 16 and reported in Table 6. Again, one can see that the indeces $n(U)$ and $n(B)$ are larger than the values obtained for the $VRI$ bands (which are closee to 2).

For comparison, we overplot the $UBVRI$-band fluxes of some SNe~Ia with very early-time observations in the right panel of Figure 16, such as SN 2009ig (Foley et al. 2012), SN 2012cg (Marion et al. 2015), SN 2013dy (Zheng et al. 2013), and ASASSN-14lp (Shappee et al. 2015), with the peak flux of each SN normalized to that of SN 2011fe. One can see that all of these comparison SNe~Ia clearly exhibit slower rises at early times, while they also have smaller $\Delta m_{15}(B)$ than SN 2011fe. However, comparison of the $\Delta m_{15}(B)$-corrected rise times between different SNe~Ia may not make much sense, given that the duration of the very early light curve does not correlate with the light-curve shape and the $n$ index shows a large range for the $t^{n}$ evolution of different SNe~Ia (Firth et al. 2015). Moreover, SN 2009ig and ASASSN-14lp have large photospheric velocities around the time of maximum light (e.g., $v_{\rm Si} \approx 13,000$ km s$^{-1}$), and these rapidly expanding SNe~Ia usually have shorter rise times compared to their normal counterparts (Zhang et al. 2010; Ganeshalingam et al. 2011); also, the early light curve of SN 2012cg may be affected by interaction with the companion star (Marion et al. 2015). On the other hand, since SN 2011by is considered to be a ``twin'' of SN 2011fe (Graham et al. 2015b), it is interesting to include it in Figure 16 even though its light curve does not cover the very early phases. As can be seen, SN 2011by is brighter than SN 2011fe at $t \approx -12.5$ days, especially in the $B$ band (see also the residual plot of Figure 16). This likely suggests that SN 2011fe may have an intrinsically faster rise rate compared to SN 2011by, since these two SNe~Ia have very similar $\Delta m_{15}(B)$ values, consistent with the argument that the gamma-ray photons from the radioactive decay are less trapped in SN 2011fe and reach its photosphere more easily relative to SN 2011by.

\subsection{The Late-Time Light Curves}

The late-time light curves can be used to constrain the underlying physics for the lingering light, such as radioactive decay of long-lived isotopes (Milne et al. 1999, 2001), interaction with CSM, and light echoes (Li et al. 2002).

Very late-time observations are rare for SNe~Ia. SN 2011fe provides us a good opportunity to study the late-time evolution of a normal SN~Ia. Based on late-time mid-IR photometry and nebular spectra, McClelland et al. (2013) found that the singly ionized iron-peak elements faded at close to the $^{56}$Co radioactive decay rate, while doubly ionized cobalt faded at a rate more than twice the $^{56}$Co radioactive decay rate owing to recombination. In Section 3.1, we noticed that SN 2011fe showed an apparently faster decay rate in the $U$ band compared to SN 2003du and SN 2003hv, which can be also interpreted as an opacity effect. Alternatively, the SN has a relatively clean environment and the expanding ejecta do not interact with the CSM expelled from the progenitor.

To obtain better knowledge of the late-time emission, we use our optical light curves to construct the late-time bolometric light curve by assuming that the contribution of NIR-band emission is about 5\% and the contribution of the UV-band emission is negligible after $t \approx +80$ days, as adopted in the analysis of SN 2003hv (Leloudas et al. 2009). We fit the bolometric light curve during the phase from $t = +80$ days to $t = +250$ days using a simple model, $L = 1.3 \times M_{\rm Ni}^{-t/111.3}(1-0.966e^{-\tau})$, where $L$ is the bolometric luminosity, $M_{\rm Ni}$ is the $^{56}$Ni mass, $\tau=(t_1/t)^2$ is the optical depth, and $t_1$ is the time when the optical depth to the gamma rays becomes unity (e.g., Sollerman et al. 1998; Leloudas et al. 2009), as shown in the subset of Figure 7. We derive $t_1=34.5$ days and $M_{\rm Ni}=0.32$~M$_\odot$, which is lower than the $^{56}$Ni mass obtained with the peak luminosity ($\sim0.57$~M$_{\odot}$). This large difference is perhaps caused by a substantial fraction of the flux being emitted beyond the UV through IR bands, or by positron escape and/or an IR catastrophe (IRC) that occurs in the ejecta at very late phases as suggested by Leloudas et al. (2009). The effect of an IRC in SN 2011fe has been confirmed by Fransson \& Jerkstrand (2015) using the spectrum taken at $\sim 1000$ days. In comparison, Mazzali et al. (2015) derived the $^{56}$Ni mass as $\sim0.47\pm0.05$~M$_{\odot}$ and the stable iron mass as $\sim 0.23\pm0.03$~M$_{\odot}$ for SN 2011fe, based on modeling of the nebular spectra.

\subsection{Progenitor Properties of SN 2011fe}
Although the progenitor properties of SN 2011fe have been thoroughly studied in the literature, our extensive observations presented here still enable us to put useful constraints from a different perspective because the UV- or $U$-band emission may be sensitive to metallicity (e.g., H{\"o}flich et al. 1998; Lentz et al. 2000; Sauer et al. 2008). The stronger UV emission and fast-rising evolution seen in SN 2011fe indicate that the ejecta from the exploding WD may have smaller opacity compared to that of some normal SNe~Ia (especially its ``twin'' SN 2011by; Foley \& Kirshner 2013; Graham et al. 2015b). This is further supported by the fast decay of the $U$-band light curve seen at late times, which is likely to be the result of less energy trapping in the ejecta.

The unusual behavior of SN 2011fe shown in the UV and $U$ bands can be reasonably explained if the progenitor star of SN 2011fe has a lower metallicity than normal. This conclusion is supported by some other evidence and analysis. Assuming that SN 2011fe is in the plane of the galactic disk of M101, and adopting a gas-phase oxygen abundance gradient of Bresolin et al. (2007), Stoll et al. (2011) estimated that the oxygen abundance inferred at the site of SN 2011fe is 12 + log(O/H) $= 8.45 \pm 0.05$, which is apparently lower than the corresponding solar value. Analysis of the observed UV spectra of SN 2011fe and SN 2011by also suggests that the former has a subsolar progenitor metallicity (Foley \& Kirshner 2013; however, see Graham et al. 2015b). The effect of metallicity variations in the progenitor on the synthetic spectra was recently studied by Baron et al. (2015), who found that a delayed-detonation model with a progenitor metallicity of Z$_\odot$/20 can fit the spectra of SN 2011fe better than a metallicity of Z$_\odot$.

The sign of a metallicity effect on the explosions of SNe~Ia has been controversial (e.g., Timmes et al. 2003; Brown et al. 2015; Miles et al. 2015). However, the study of SN 2011fe seems to indicate that the emission of SNe~Ia at shorter wavelengths could be enhanced with decreasing metallicity of the progenitors.

\section{CONCLUSIONS}

In this paper, we present extensive observations of optical photometry and spectroscopy of SN 2011fe. Fitting to the early-time multi-band light curves of SN 2011fe with a power-law model ($f\propto(t-t_{0})^{n}$) yields a rise time of 18.00 days, which is consistent with previous results. The best-fit index $n$ is close to 2 in the $VRI$ bands, but larger than 2 in the $UB$ bands, suggestive of a faster rise at shorter wavelengths. SN 2011fe is also found to have stronger UV emission and reach its UV peak a few days earlier than the comparison SNe~Ia such as SN 2005cf and the ``twin'' SN 2011by. Moreover, the $U$-band light curve shows a remarkably faster decay rate at late times compared to other normal SNe~Ia.

The early-time spectra of SN 2011fe resemble those of other normal SNe~Ia in many respects, including the presence of C~II absorption and the HVFs from the Ca~II~NIR triplet. Note that no significant HVF is detected in Si~II $\lambda$6355 absorption even in the extremely early-time spectrum. The HVFs are relatively weak in SN 2011fe compared to other normal SNe~Ia with similar $\Delta m_{15}(B)$. On the other hand, the O-HVFs are very prominent in SN 2011fe, strong relative to normal SNe~Ia. Moreover, a second HVF at higher velocities ($\sim$ 22,000 km~s$^{-1}$) can be also identified in some of our early-time spectra. The presence of this high-velocity oxygen indicates that the burning of the C+O white dwarf is not complete for SN 2011fe and the corresponding photosphere has a lower temperature. This is consistent with the relatively weak Si-HVF in SN 2011fe, given an ionization (or temperature) effect and/or an abundance enhancement scenario for the formation of HVFs.

These results suggest congruously that the expanding ejecta of the progenitor of SN 2011fe have a lower opacity and hence a lower metallicity. This conclusion is consistent with a recent result obtained through modeling of early-time spectra of SN 2011fe (Baron et al. 2015).

\acknowledgments
We are grateful to the staffs of the various telescopes and observatories with which data were obtained (Tsinghua-NAOC Telescope, Lijiang Telescope, Xinglong 2.16~m Telescope, KAIT, Yunnan Astronomical Observatory, Lick Observatory). Financial support for this work has been provided by the National Science Foundation of China (NSFC grants 11178003, 11325313, 11403096, 11203034), the Major State Basic Research Development Program (2013CB834903) and the Key Research Program of the Chinese Academy of Sciences (Grant NO. KJZD-EW-M06). This work was partially supported by the Open Project Program of the Key Laboratory of Optical Astronomy, National Astronomical Observatories, Chinese Academy of Sciences. A.V.F.'s group at UC Berkeley is grateful for financial assistance from NSF grant AST-1211916, the TABASGO Foundation, and the Christopher R. Redlich Fund. Research at Lick Observatory is partially supported by a generous gift from Google. KAIT and its ongoing operation were made possible by donations from Sun Microsystems, Inc., the Hewlett-Packard Company, AutoScope Corporation, Lick Observatory, the NSF, the University of California, the Sylvia \& Jim Katzman Foundation, and the TABASGO Foundation. Funding for the LJT has been provided by the Chinese Academy of Sciences and the People's Government of Yunnan Province.

{}

\clearpage

\begin{figure}[htbp]
\center
\includegraphics[angle=0,width=1\textwidth]{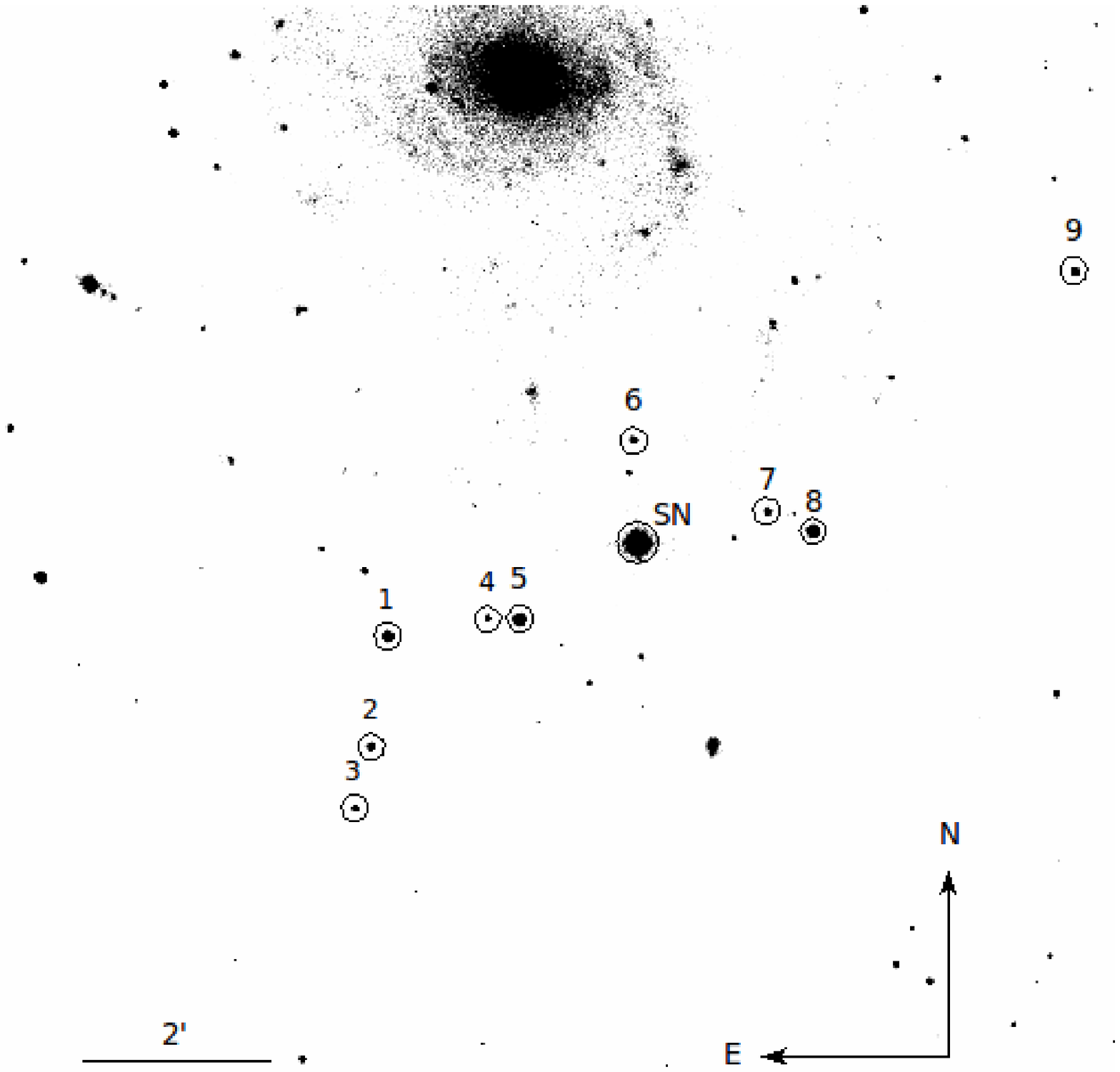}
\vspace{0.2cm}
\caption{SN 2011fe in M101. This is a $R$-band image taken with the TNT 0.8~m telescope on 2011 Sep. 20. The supernova and 9 local reference stars are marked. North is up and east is to the left.}
\label{fig-1} \vspace{-0.0cm}
\end{figure}

\clearpage
\begin{figure*}[htbp]
\center
\includegraphics[angle=0,width=1\textwidth]{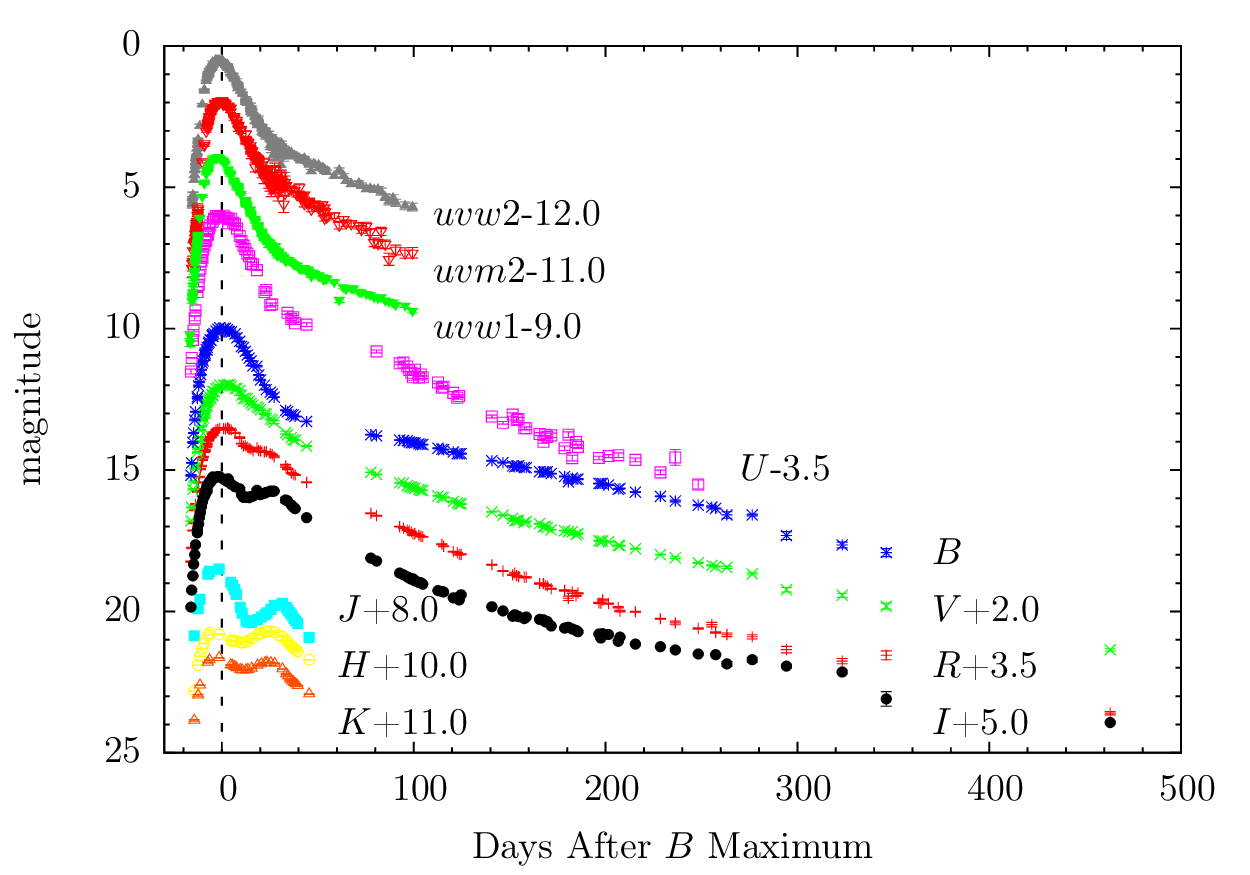}
\vspace{0.2cm}
\caption{The UV, optical, and NIR light curves of SN 2011fe, with UV data from Brown et al. (2012) and NIR data from Matheson et al. (2012).}
\label{fig-2} \vspace{-0.0cm}
\end{figure*}

\clearpage
\begin{figure}[htbp]
\center
\includegraphics[angle=0,width=1\textwidth]{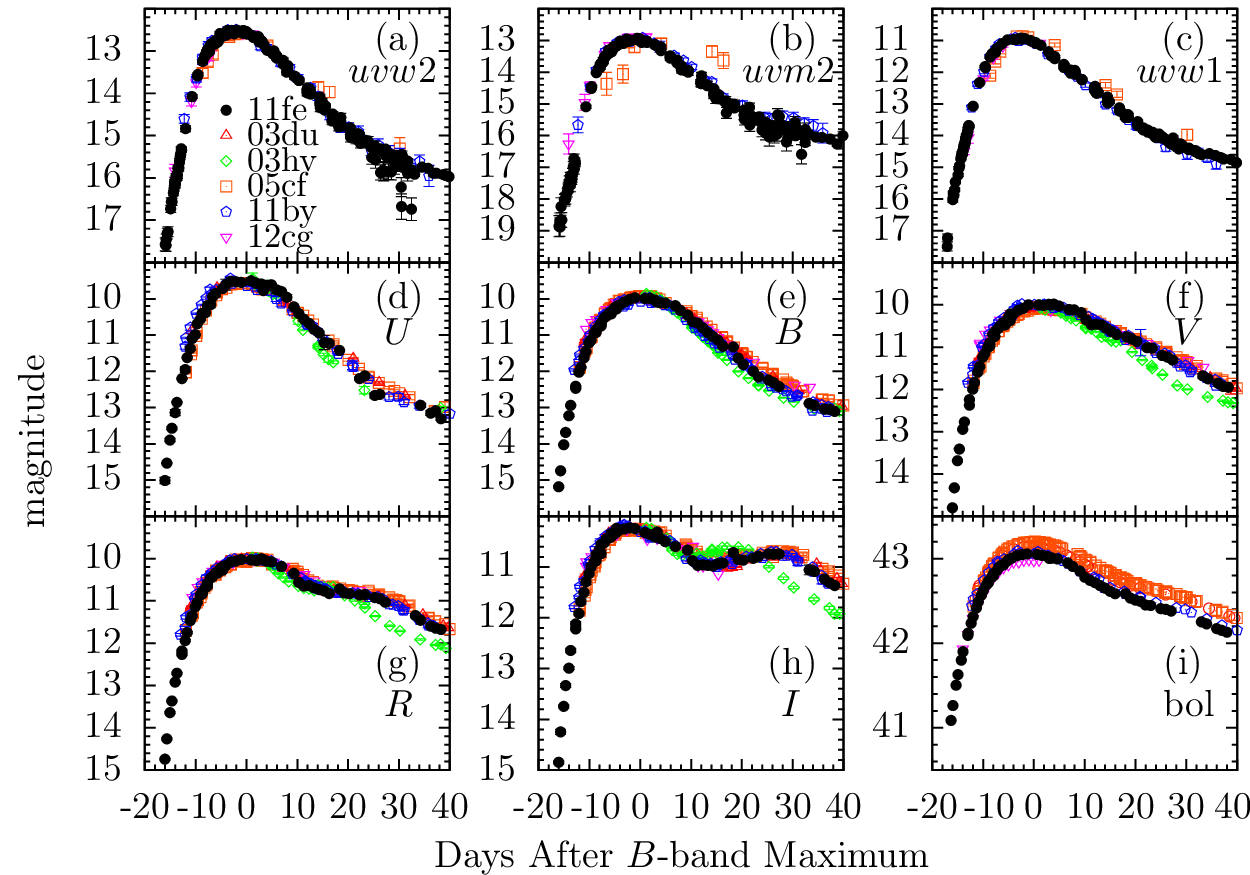}
\vspace{0.2cm}
\caption{Comparison of the near-maximum-light UV, optical, and bolometric (erg s$^{-1}$) light curves of SN 2011fe and other well-observed SNe~Ia: SN 2003du, SN 2003hv, SN 2005cf, SN 2011by, and SN 2012cg. See text for references.}
\label{fig-3} \vspace{-0.0cm}
\end{figure}

\clearpage
\begin{figure}[htbp]
\center
\includegraphics[angle=0,width=1\textwidth]{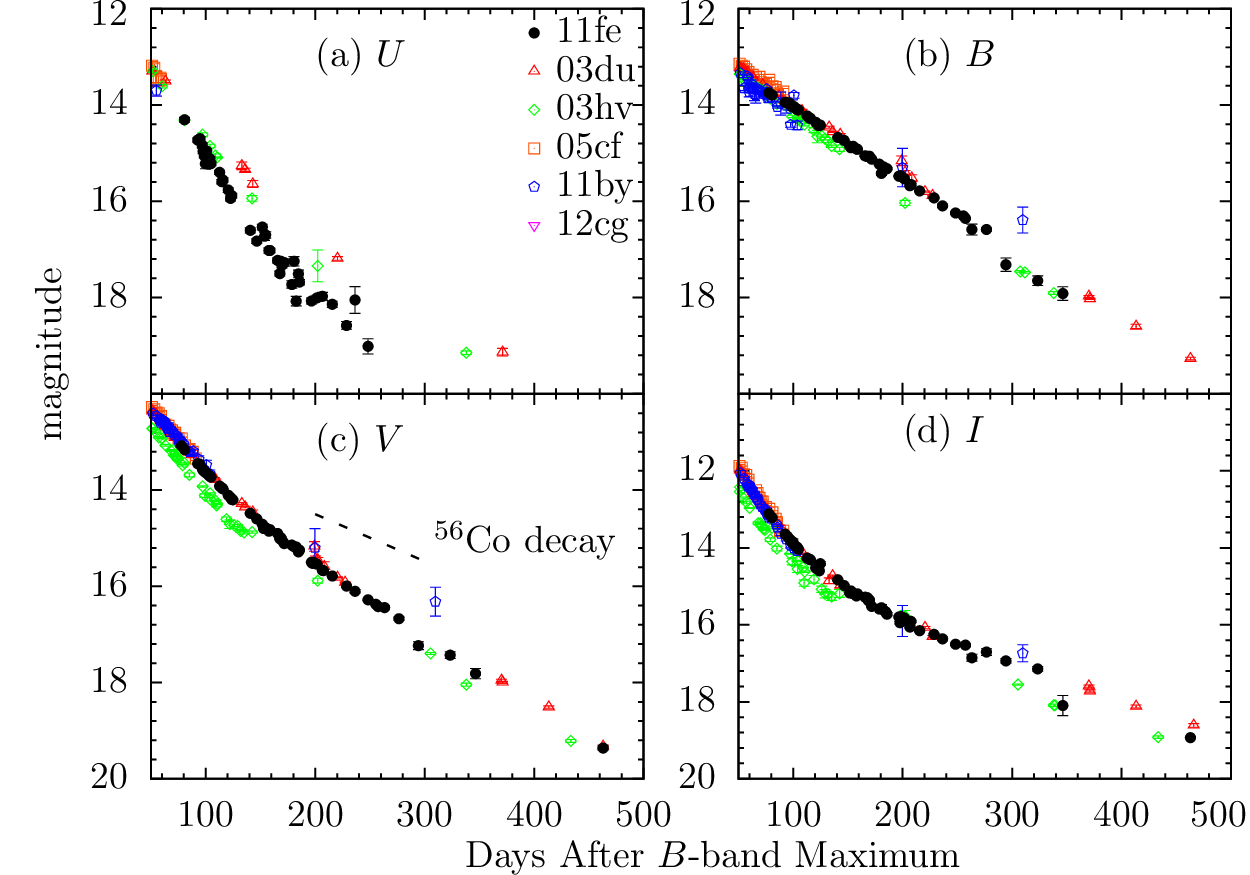}
\vspace{0.2cm}
\caption{Comparison of the late-time $UBVI$  light curves of SN 2011fe and other well-observed SNe~Ia: SN 2003du, SN 2003hv, SN 2005cf, SN 2011by, and SN 2012cg. The $^{56}$Co $\rightarrow$ $^{56}$Fe decay rate is also plotted. See text for references.}
\label{fig-4} \vspace{-0.0cm}
\end{figure}

\clearpage
\begin{figure}[htbp]
\center
\includegraphics[angle=0,width=1\textwidth]{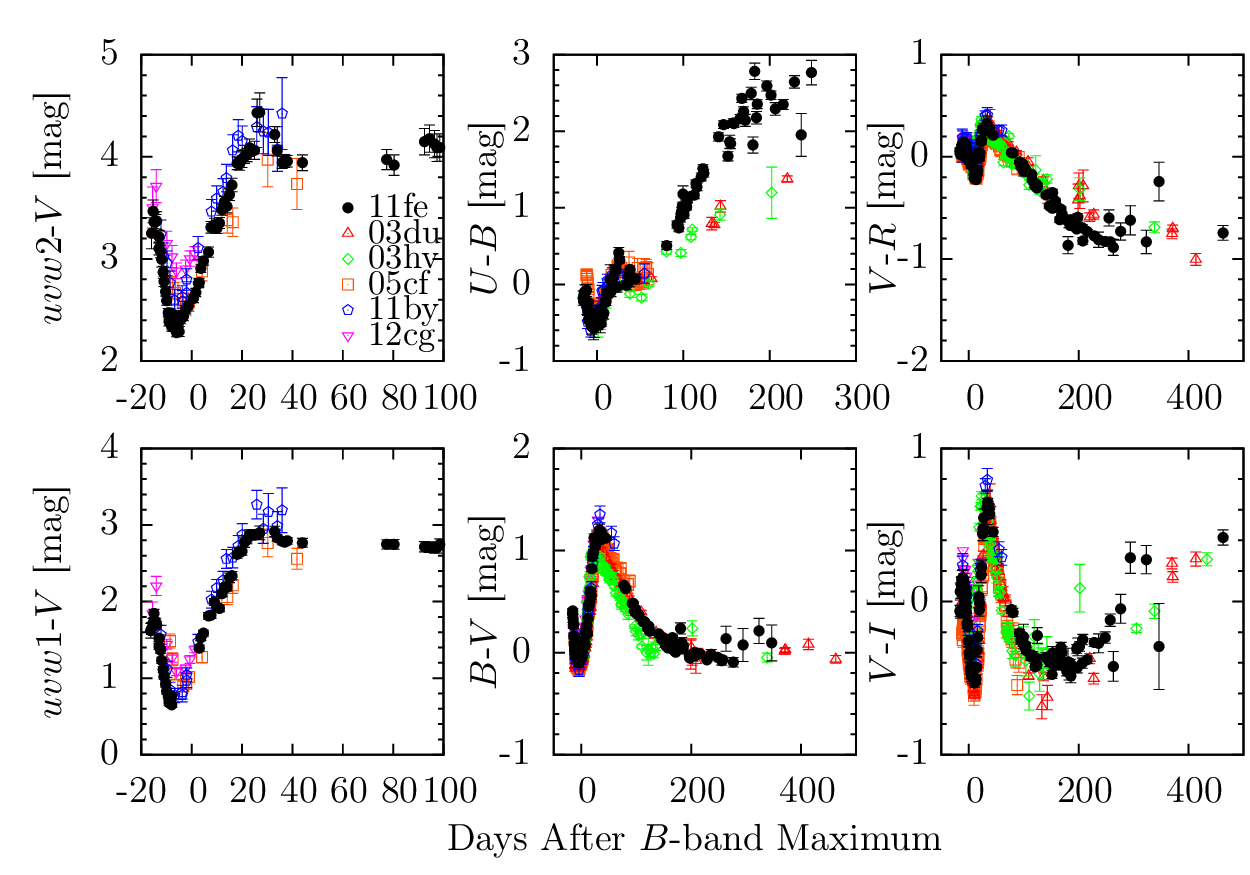}
\vspace{0.2cm}
\caption{$uvw2-V$, $uvw1-V$, $U-B$, $B-V$, $V-R$, and $V-I$ color curves of SN 2011fe compared with those of SNe 2003du, 2003hv, 2005cf, 2011by, and 2012cg. All of the color curves have been corrected for reddening from the Milky Way and the host galaxies. The data sources are cited in the text. The symbols are the same as in Fig. 3.}
\label{fig-5} \vspace{-0.0cm}
\end{figure}

\clearpage
\begin{figure}[htbp]
\center
\includegraphics[angle=0,width=1\textwidth]{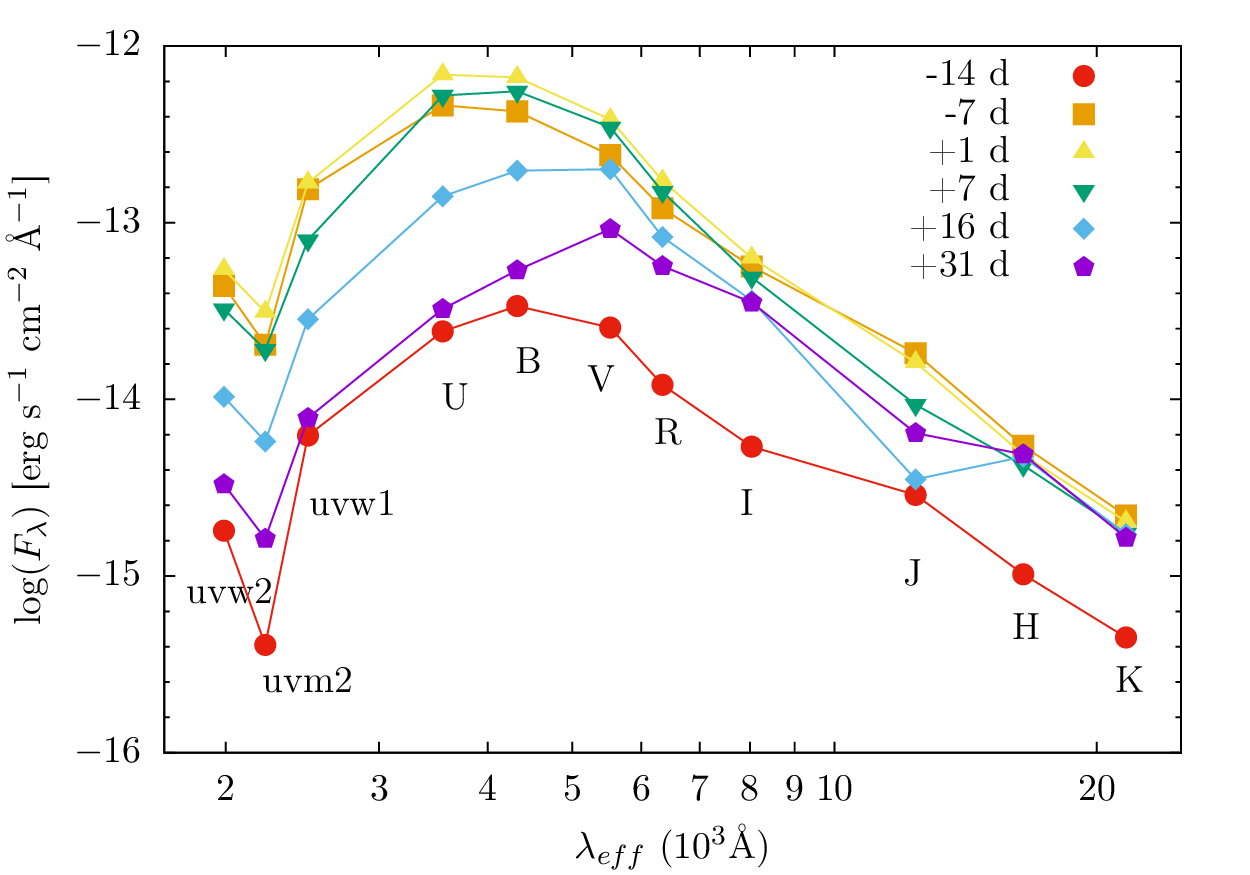}
\vspace{0.2cm}
\caption{The spectral energy distribution of SN 2011fe obtained at $t \approx -14$, $-7$, +1, +7, +16, and +31 days after $B$-band maximum.}
\label{fig-6} \vspace{-0.0cm}
\end{figure}

\clearpage
\begin{figure}[htbp]
\center
\includegraphics[angle=0,width=1\textwidth]{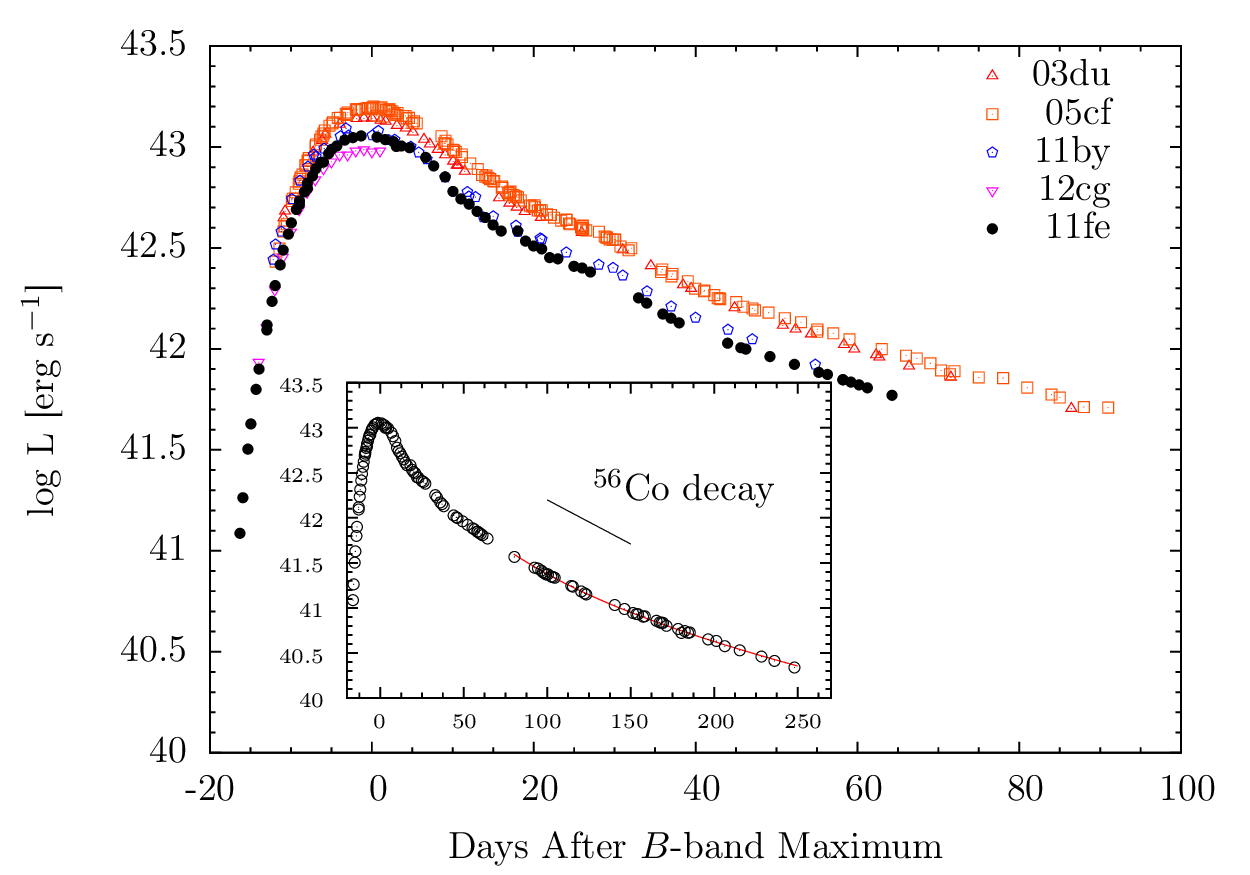}
\vspace{0.2cm}
\caption{The UV through optical bolometric light curve of SN 2011fe compared with those of SN 2003du, SN 2005cf, SN 2011by, and SN 2012cg. The late-time bolometric light curve of SN 2011fe is shown in the inset. The red line is the model described in the text, and the black line shows the $^{56}$Co $\rightarrow$ $^{56}$Fe decay.}
\label{fig-7} \vspace{-0.0cm}
\end{figure}

\clearpage
\begin{figure}[htbp]
\center
\includegraphics[angle=0,width=1\textwidth]{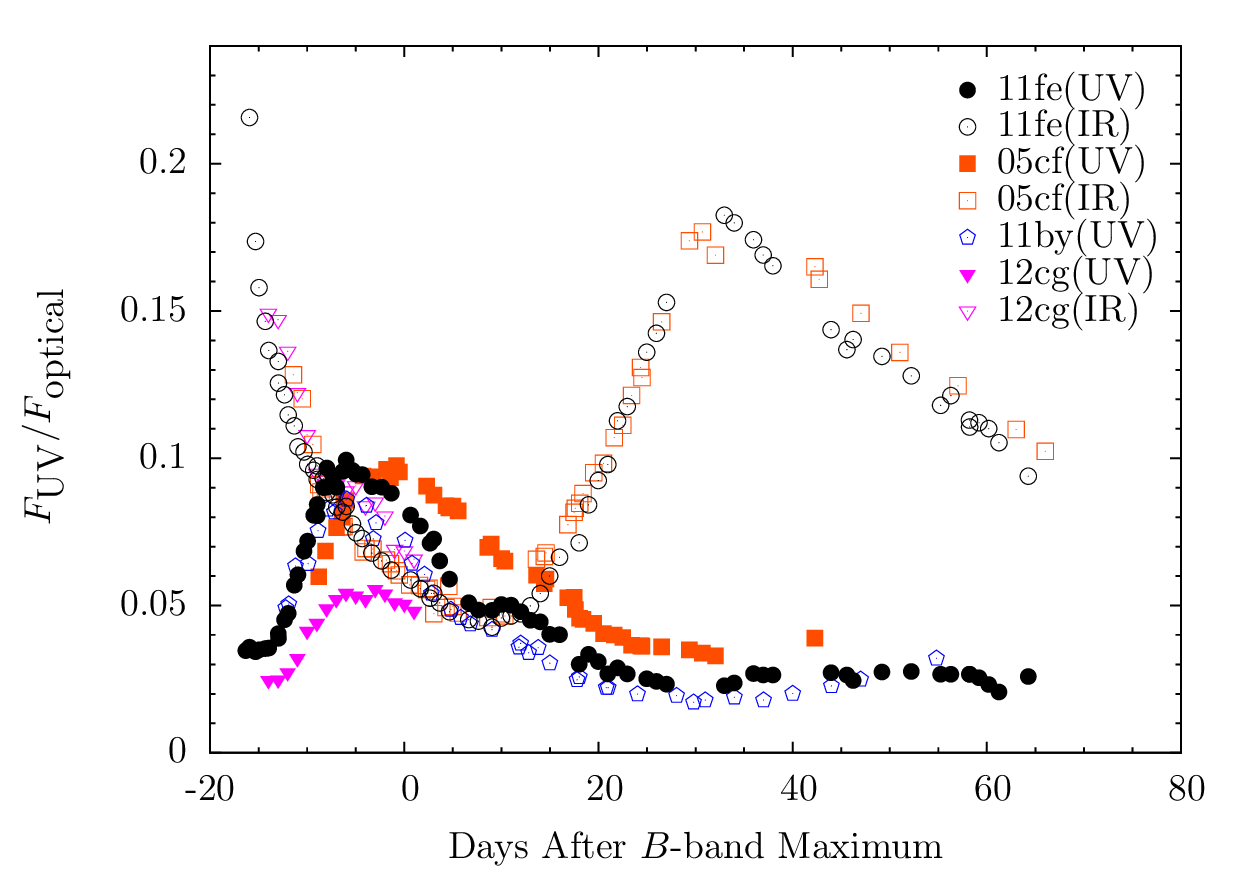}
\vspace{0.2cm}
\caption{The ratio of the UV (1600--3200~\AA) and NIR (9000--24,000~\AA) fluxes to the optical (3200--9000~\AA). Overplotted are the corresponding flux ratios of SN 2005cf, SN 2011by, and SN 2012cg.}
\label{fig-8} \vspace{-0.0cm}
\end{figure}

\clearpage
\begin{figure}[htbp]
\center
\includegraphics[angle=0,width=1\textwidth]{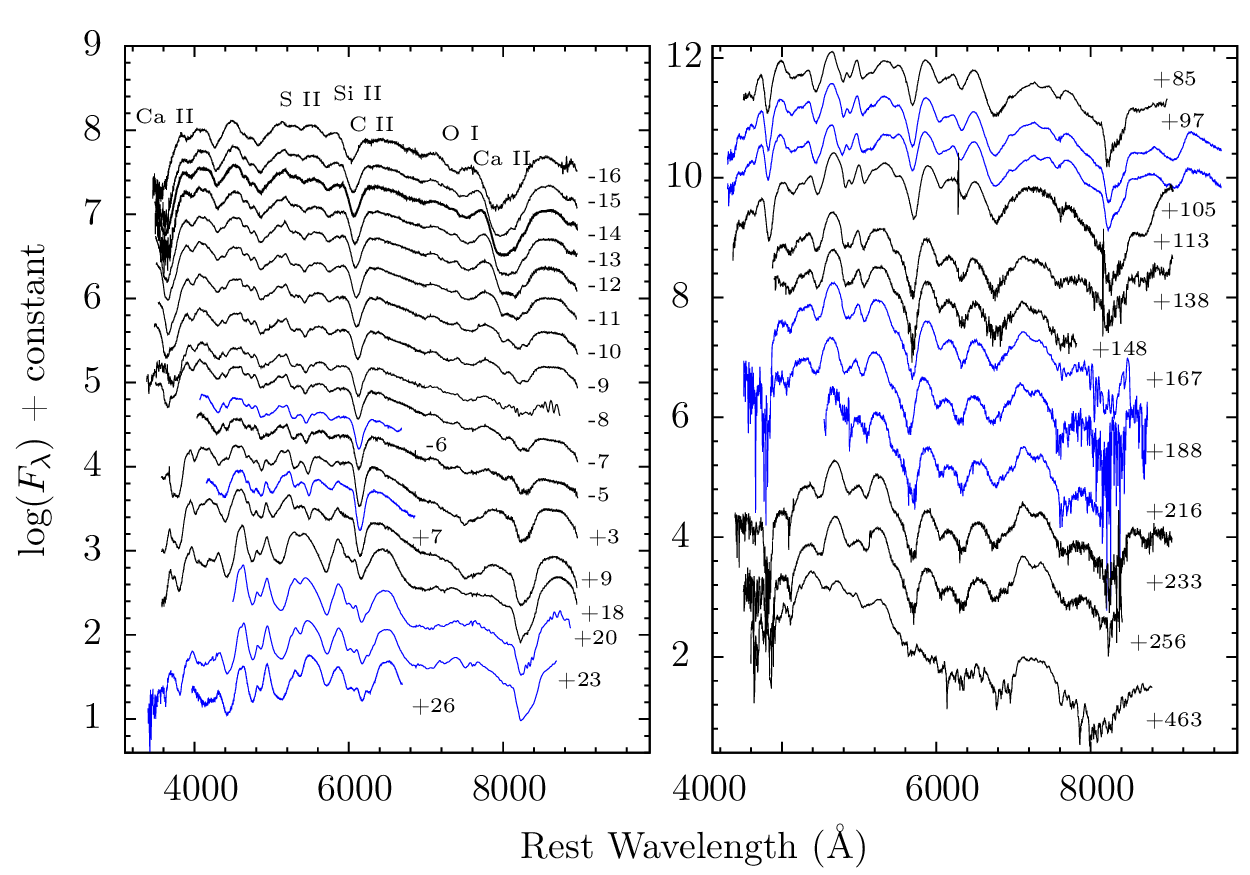}
\vspace{0.cm}
\caption{Optical spectra of SN 2011fe. Spectra obtained with the YNAO 2.4~m telescope are shown in black; those obtained with the Xinglong 2.16~m of NAOC are in blue.}
\label{fig-9} \vspace{-0.0cm}
\end{figure}

\clearpage
\begin{figure}[htbp]
\center
\includegraphics[angle=0,width=1\textwidth]{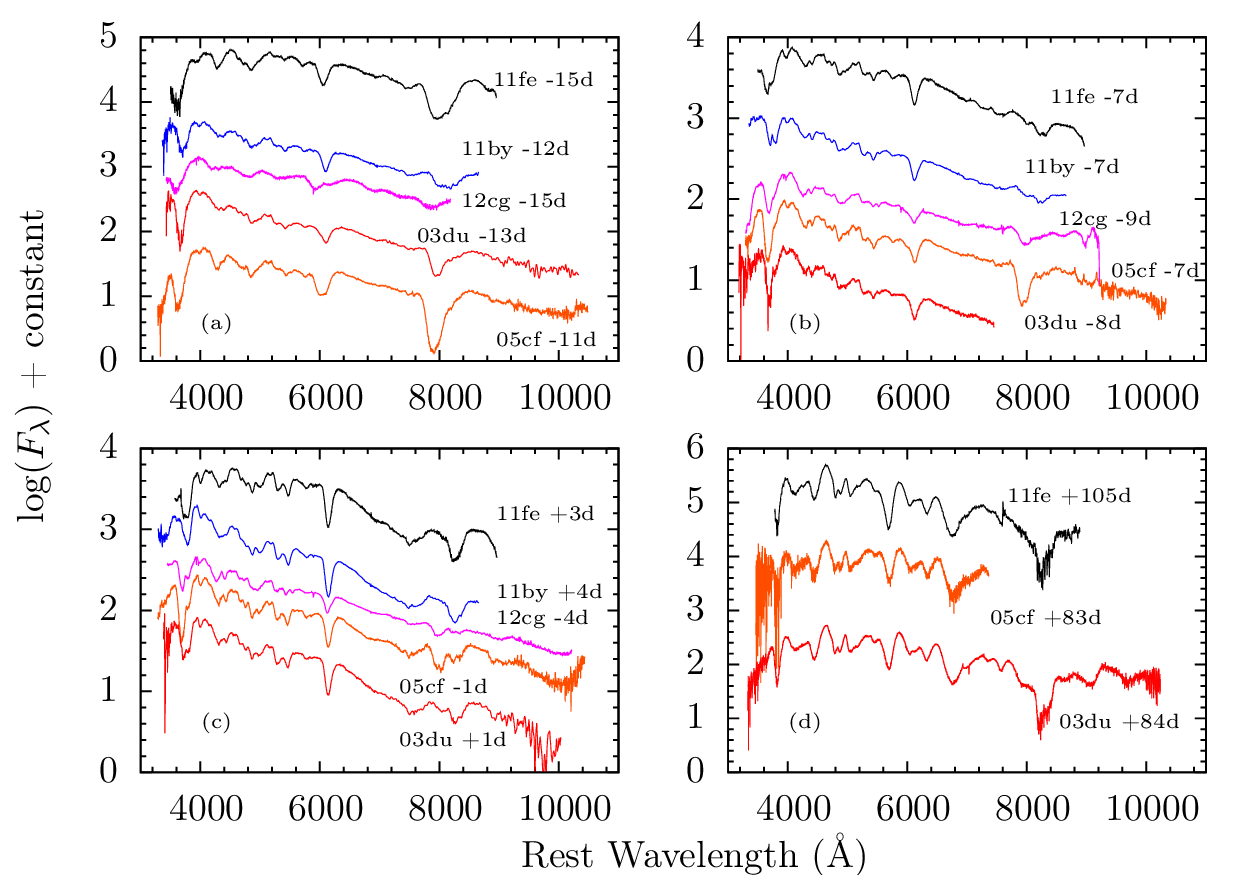}
\vspace{0.2cm}
\caption{The spectrum of SN 2011fe at $t \approx -15$ d, $-7$ d, 0 d, and +3 months after $B$-band maximum. The comparable-phase spectra of SN 2003du, SN 2005cf, SN 2011by, and SN 2012cg are overplotted for comparison.}
\label{fig-10} \vspace{-0.0cm}
\end{figure}

\clearpage
\begin{figure}[htbp]
\center
\includegraphics[angle=0,width=1\textwidth]{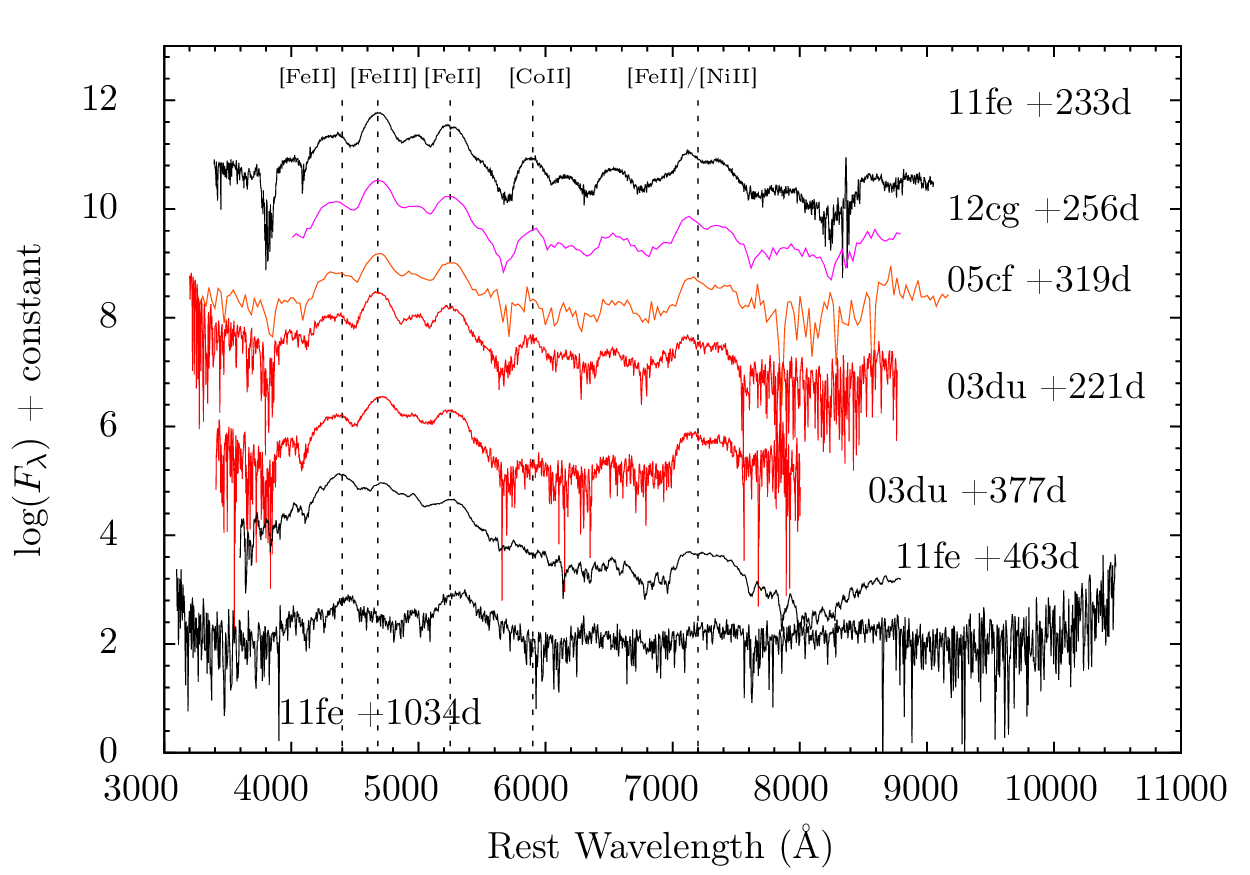}
\vspace{0.2cm}
\caption{Late-time spectra of SN 2011fe. The nebular spectra of SN 2003du, SN 2005cf, SN 2011by, and SN 2012cg at similar phases are also shown for comparison.}
\label{fig-11} \vspace{-0.0cm}
\end{figure}

\clearpage
\begin{figure}[htbp]
\center
\includegraphics[angle=0,width=1\textwidth]{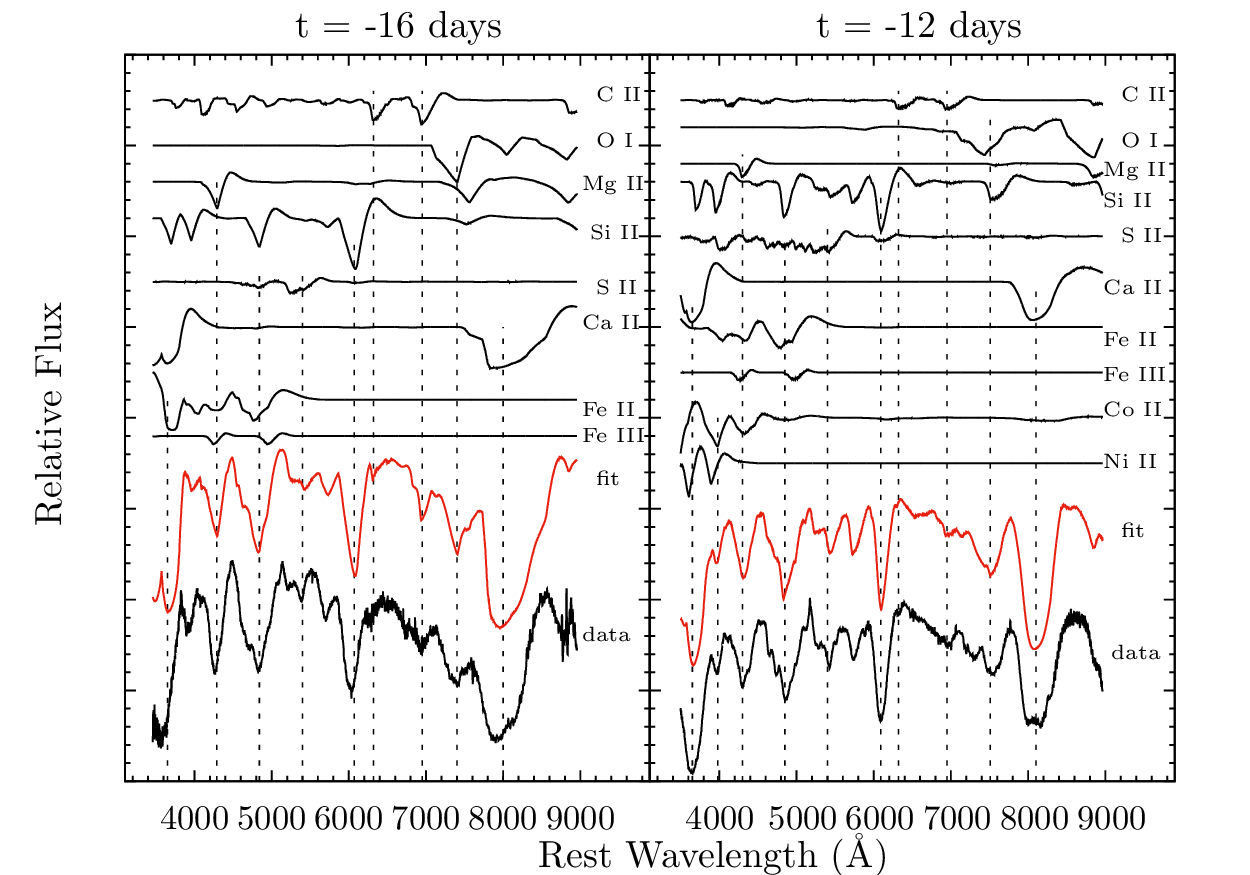}
\vspace{0.2cm}
\caption{SYNAPPS fit to the $t = -16$ d and $-12$ d spectra of SN 2011fe. Features from different ions are marked with dashed lines.}
\label{fig-12} \vspace{-0.0cm}
\end{figure}

\clearpage
\begin{figure}[htbp]
\center
\includegraphics[angle=0,width=1\textwidth]{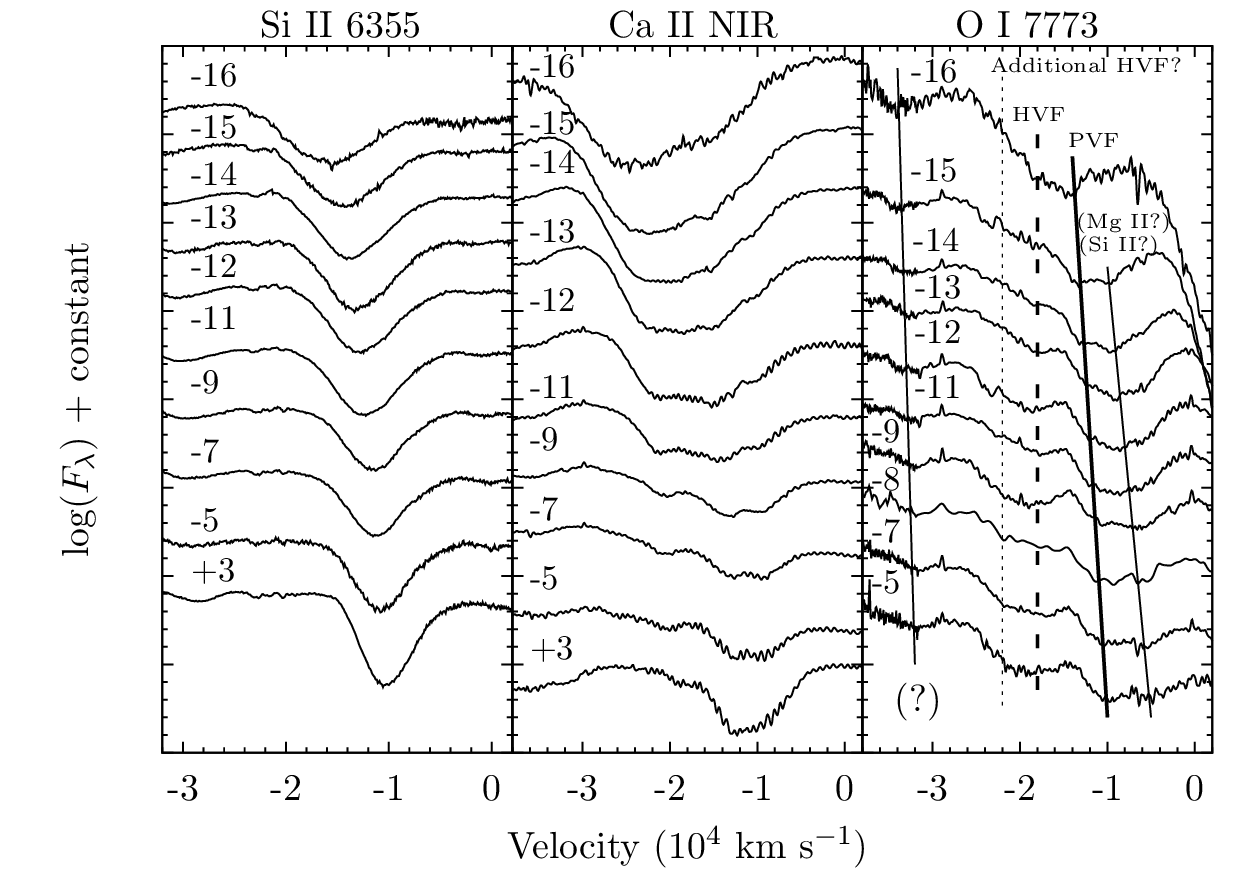}
\vspace{0.2cm}
\caption{Evolution of Si~II $\lambda$6355, Ca~II~NIR, and O~I $\lambda$7773 lines of SN 2011fe in velocity space. The solid line marks the positions of the photospheric component, the dashed line marks the high-velocity feature, and the dotted line marks the position of a possible additional HVF of O~I $\lambda$7773 identified by Zhao et al. (2016). }
\label{fig-13} \vspace{-0.0cm}
\end{figure}

\clearpage
\begin{figure}[htbp]
\center
\includegraphics[angle=0,width=1\textwidth]{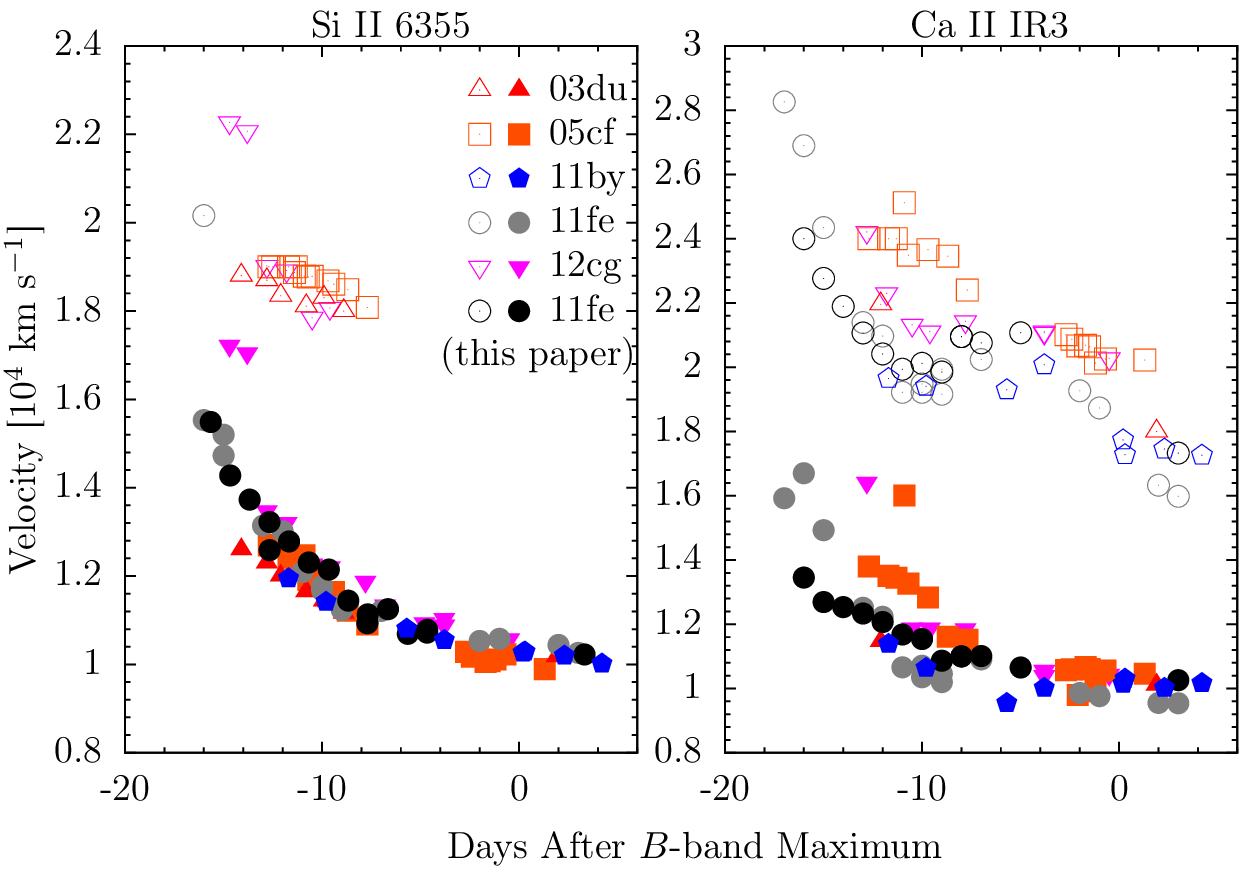}
\vspace{0.2cm}
\caption{Evolution of the expansion velocity of SN 2011fe as measured from the absorption minima of Si~II $\lambda$6355 (left panel) and the Ca~II~NIR triplet (right panel), in comparison with the values of SN 2003du, SN 2005cf, SN 2011by, and SN 2012cg (see text for references). Open symbols represent the HVFs and filled symbols represent the PVFs.}
\label{fig-14} \vspace{-0.0cm}
\end{figure}

\clearpage
\begin{figure}[htbp]
\center
\includegraphics[angle=0,width=1\textwidth]{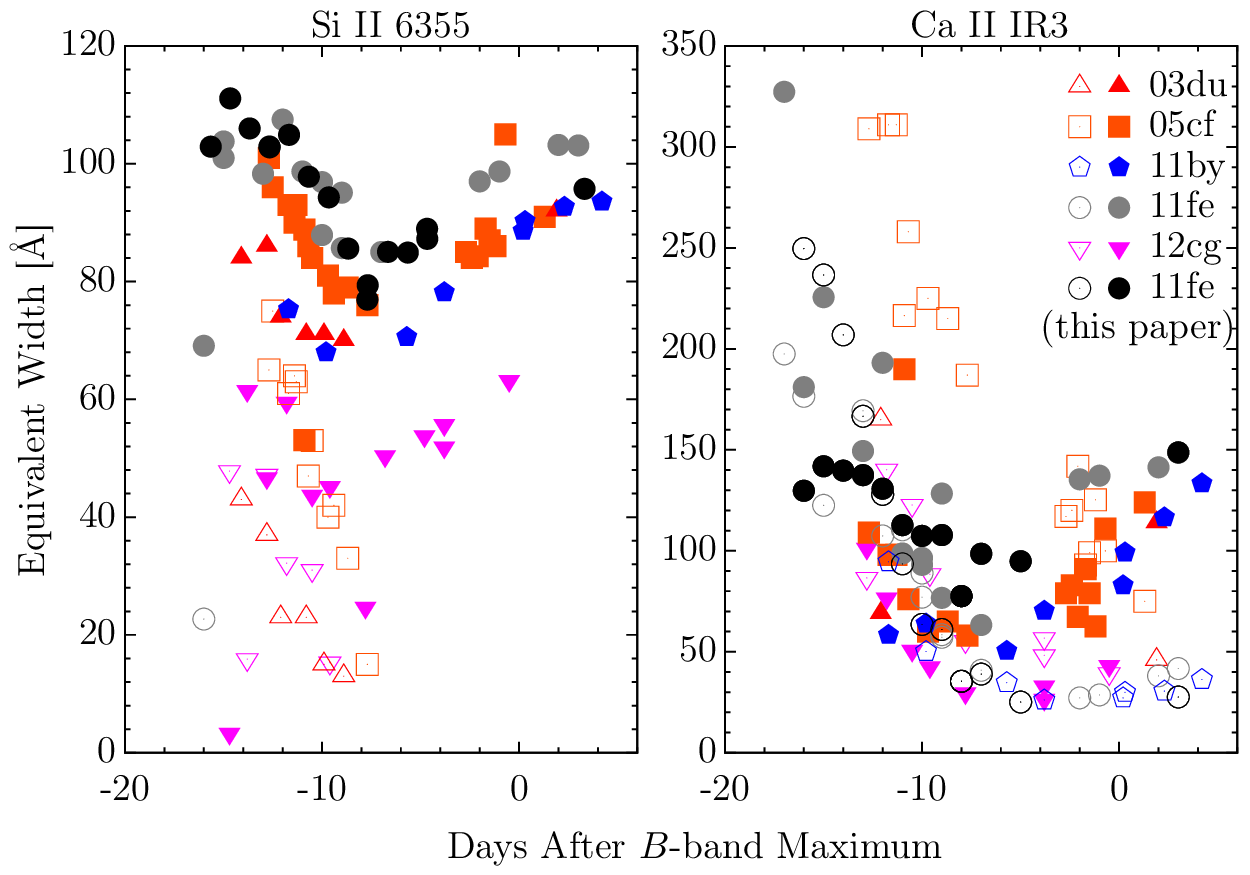}
\vspace{0.2cm}
\caption{Evolution of the pseudo-equivalent width (pEW) of SN 2011fe as measured from the absorption of Si~II $\lambda$6355 (left panel) and the Ca~II~NIR triplet (right panel), compared with the measurements of SN 2003du, SN 2005cf, SN 2011by, and SN 2012cg (see text for references). Open symbols denote the HVFs and filled symbols denote the PVFs.}
\label{fig-15} \vspace{-0.0cm}
\end{figure}

\clearpage
\begin{figure}[htbp]
\center
\includegraphics[angle=0,width=1\textwidth]{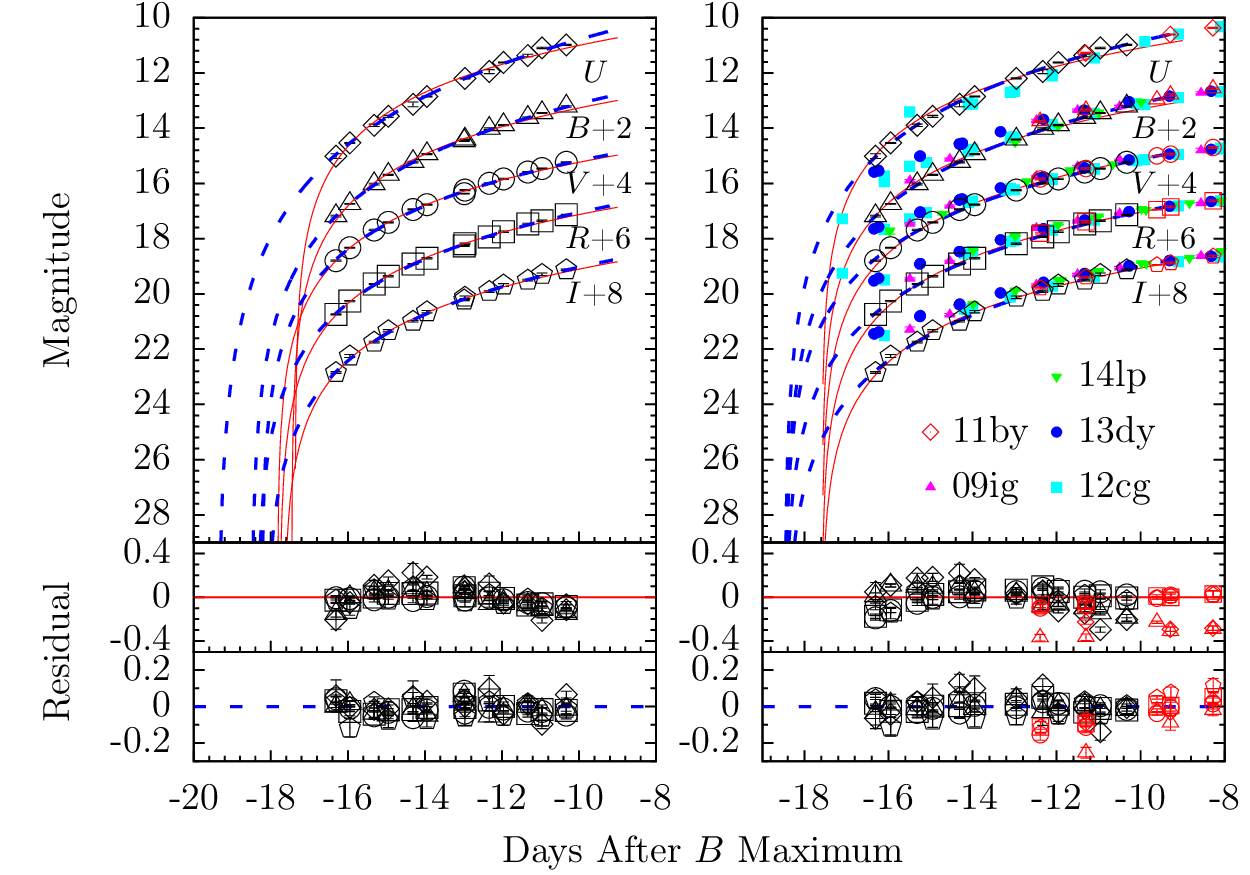}
\vspace{0.2cm}
\caption{Fit to the early-time $UBVRI$ light curves of SN 2011fe. Left panel shows fits to the data with $t < -10$ days; the red line shows the fit with the $(t-t_{\rm first})^2$ model, while the blue line represents the fit with the $(t-t_0)^n$ model. Right panel shows a multi-band $t^n$ fitting by forcing all five bands to have the same $t_{\rm first}$; the red line shows the fit with the $(t-t_{\rm first})^2$ model, while the blue line represents the fit with the $(t-t_0)^n$ model. Some SNe~Ia with early-time data are also plotted for comparison.}
\label{fig-16} \vspace{-0.0cm}
\end{figure}

\begin{deluxetable}{cccccccc}
\tablecolumns{6} \tablewidth{0pc} \tabletypesize{\scriptsize}
\tablecaption{Photometric Standards in the SN 2011fe
Field\tablenotemark{a}} \tablehead{\colhead{Star} &
\colhead{$\alpha$(J2000)} & \colhead{$\delta$(J2000)} & \colhead{$U$
(mag\tablenotemark{b})} & \colhead{$B$ (mag)} & \colhead{$V$ (mag)}
& \colhead{$R$ (mag)} & \colhead{$I$ (mag)} } \startdata
1& 14$^h$03$^m$22.39$^s$ & +$54^\circ15'35.9''$ & 17.369(139) & 16.142(023)& 14.861(015)& 14.007(008)&13.290(011) \\
2& 14:03:23.75 & +54:14:32.6 & 16.023(111) & 16.077(019)& 15.452(017)& 15.089(012)&14.700(011) \\
3& 14:03:24.91 & +54:13:57.2 & 16.840(020) & 16.858(007)& 16.151(009)& 15.877(013)&15.499(018) \\
4& 14:03:15.81 & +54:15:44.8 & 17.636(120) & 17.165(027)& 16.487(014)& 16.116(018)&15.707(016) \\
5& 14:03:13.66 & +54:15:43.3 & 15.111(085) & 14.778(020)& 13.849(010)& 13.319(008)&12.781(011) \\
6& 14:03:05.82 & +54:17:25.3 & 16.901(030) & 16.779(019)& 16.120(013)& 15.659(007)&15.248(008) \\
7& 14:02:57.09 & +54:16:41.0 & 16.510(072) & 16.596(024)& 16.090(007)& 15.743(005)&15.379(007) \\
8& 14:02:54.12 & +54:16:29.0 & 14.683(079) & 14.608(019)& 14.005(009)& 13.653(010)&13.296(012) \\
9& 14:02:36.29 & +54:18:54.3 & 18.287(100) & 17.027(038)& 15.663(035)& 14.750(023)&13.985(021)
\enddata
\tablenotetext{a}{See Figure 1 for a chart of SN 2011fe and the comparison stars.}
\tablenotetext{b}{Note: Uncertainties, in units of 0.001 mag, are $1\sigma$.}
\end{deluxetable}

\begin{deluxetable}{cccccccc}
\tablecolumns{6} \tablewidth{0pc} \tabletypesize{\scriptsize}
\tablecaption{Optical Photometry of SN 2011fe}
\tablehead{\colhead{JD\tablenotemark{a}} &
\colhead{Phase\tablenotemark{b}} &\colhead{$U$
(mag)\tablenotemark{c}} &\colhead{$B$ (mag)} &\colhead{$V$ (mag)}
&\colhead{$R$ (mag)} &\colhead{$I$ (mag)}  & \colhead{Telescope}}
\startdata
5798.17 & -16.31 & 15.013(086) & 15.185(030) & 14.797(021) & 14.740(020) & 14.848(030) & KAIT   \\
5798.53 & -15.95 & 14.534(014) & 14.743(013) & 14.325(019) & 14.260(024) & 14.248(052) & LJT    \\
5799.16 & -15.32 & 13.894(044) & 14.027(030) & 13.685(020) & 13.641(020) & 13.745(030) & KAIT   \\
5799.53 & -14.95 & 13.572(036) & 13.687(019) & 13.410(020) & 13.367(025) & 13.333(040) & LJT    \\
5800.17 & -14.31 & 13.138(084) & 13.225(030) & 12.943(020) & 12.921(020) & 12.991(030) & KAIT   \\
5800.53 & -13.95 & 12.855(017) & 12.940(019) & 12.768(020) & 12.710(025) & 12.645(041) & LJT    \\
5801.50 & -12.98 & \nd & 12.470(012) & 12.370(011) & 12.266(017) & 12.221(023) & TNT    \\
5801.52 & -12.96 & 12.201(017) & 12.410(017) & 12.242(021) & 12.183(025) & 12.121(046) & LJT    \\
5802.15 & -12.33 & 11.951(072) & 12.022(030) & 11.996(020) & 11.938(020) & 11.914(030) & KAIT   \\
5802.52 & -11.96 & 11.621(019) & 11.890(017) & 11.837(022) & 11.748(024) & 11.678(051) & LJT    \\
5803.15 & -11.33 & 11.371(046) & 11.624(030) & 11.590(020) & 11.464(020) & 11.500(030) & KAIT   \\
5803.52 & -10.96 & 11.094(024) & 11.454(015) & 11.460(020) & 11.353(025) & 11.293(048) & LJT    \\
5804.15 & -10.33 & 10.988(019) & 11.216(030) & 11.236(021) & 11.136(020) & 11.129(030) & KAIT   \\
5804.53 & -9.95 & 10.688(019) & 11.142(016) & 11.152(022) & 11.039(024) & 10.995(059) & LJT \\
5805.15 & -9.33 & 10.557(015) & 10.943(030) & 11.002(020) & 10.854(020) & 10.900(030) & KAIT    \\
5805.51 & -8.97 & \nd & 10.929(011) & 10.972(021) & 10.835(014) & 10.878(023) & TNT \\
5805.52 & -8.96 & 10.400(038) & 10.861(014) & 10.903(020) & 10.788(025) & 10.763(050) & LJT \\
5806.14 & -8.34 & 10.379(017) & 10.743(030) & 10.697(020) & 10.688(020) & 10.666(030) & KAIT    \\
5806.51 & -7.97 & \nd & 10.750(010) & 10.744(013) & 10.651(035) & 10.734(066) & TNT \\
5806.52 & -7.96 & 10.190(028) & 10.662(025) & 10.628(022) & 10.581(025) & 10.553(043) & LJT \\
5807.14 & -7.34 & 10.160(088) & 10.519(030) & 10.559(020) & 10.460(020) & 10.519(030) & KAIT    \\
5807.54 & -6.94 & 9.940(018) & 10.473(021) & 10.504(023) & 10.423(029) & 10.443(044) & LJT  \\
5808.14 & -6.34 & 9.852(093) & 10.396(030) & 10.426(020) & 10.338(020) & 10.425(030) & KAIT \\
5808.50 & -5.98 & \nd & 10.412(009) & 10.475(021) & 10.347(009) & 10.403(021) & TNT \\
5808.52 & -5.96 & \nd & \nd & \nd & \nd & 10.342(060) & LJT \\
5809.14 & -5.34 & \nd & 10.252(030) & 10.323(021) & 10.226(020) & 10.346(030) & KAIT    \\
5809.52 & -4.96 & 9.714(054) & 10.214(014) & 10.216(023) & 10.185(026) & 10.240(048) & LJT  \\
5809.52 & -4.96 & \nd & 10.281(010) & 10.372(034) & 10.271(021) & 10.313(043) & TNT \\
5810.14 & -4.34 & 9.607(147) & 10.171(030) & 10.212(020) & 10.167(020) & 10.270(030) & KAIT \\
5811.14 & -3.34 & 9.516(115) & 10.082(030) & 10.124(020) & 10.079(020) & 10.251(030) & KAIT \\
5812.14 & -2.34 & 9.527(028) & 10.017(030) & 10.079(020) & 10.057(020) & 10.228(030) & KAIT \\
5813.14 & -1.34 & 9.545(088) & 9.969(030) & 9.996(020) & 10.020(020) & 10.243(030) & KAIT   \\
5815.13 & 0.65 & 9.505(092) & 9.991(030) & 9.998(020) & 10.034(020) & 10.321(030) & KAIT    \\
5816.13 & 1.65 & 9.570(014) & 9.987(030) & 10.015(021) & 10.047(020) & 10.360(030) & KAIT   \\
5817.13 & 2.65 & 9.589(070) & 10.034(030) & 9.984(020) & 10.004(020) & 10.410(030) & KAIT   \\
5817.51 & 3.03 & 9.772(013) & 10.095(028) & 9.998(020) & 10.028(025) & 10.311(042) & LJT    \\
5818.13 & 3.65 & 9.631(144) & 10.108(030) & 9.983(020) & 10.062(020) & 10.439(030) & KAIT   \\
5819.13 & 4.65 & 9.613(025) & 10.117(030) & 10.025(020) & 10.081(020) & 10.501(030) & KAIT  \\
5820.13 & 5.65 & 9.754(100) & \nd & \nd & \nd & \nd & KAIT  \\
5821.12 & 6.64 & 9.800(040) & 10.188(030) & 10.107(021) & 10.187(020) & 10.594(030) & KAIT  \\
5822.12 & 7.64 & 9.958(075) & 10.324(030) & 10.117(020) & \nd & \nd & KAIT  \\
5823.52 & 9.04 & 10.224(019) & 10.447(015) & 10.184(023) & 10.351(027) & 10.667(047) & LJT  \\
5824.49 & 10.01 & 10.404(045) & 10.627(010) & 10.356(015) & 10.564(012) & 10.875(025) & TNT \\
5825.49 & 11.01 & 10.564(043) & 10.667(010) & 10.471(017) & 10.650(010) & 10.969(026) & TNT \\
5826.49 & 12.01 & 10.676(041) & 10.800(009) & 10.456(012) & 10.636(011) & 10.956(025) & TNT \\
5827.47 & 12.99 & 10.840(051) & 10.933(011) & 10.474(017) & 10.692(014) & 10.969(020) & TNT \\
5828.48 & 14.00 & 10.939(042) & 11.039(009) & 10.558(011) & 10.731(016) & 10.978(018) & TNT \\
5829.46 & 14.98 & 11.207(194) & 11.138(008) & 10.632(037) & 10.763(028) & 10.935(022) & TNT \\
5830.47 & 15.99 & 11.228(068) & 11.311(021) & 10.701(022) & 10.827(032) & 10.923(031) & TNT \\
5832.50 & 18.02 & 11.425(013) & 11.328(026) & 10.759(032) & 10.715(026) & 10.716(064) & LJT \\
5833.46 & 18.98 & \nd & 11.637(012) & 10.807(025) & 10.814(012) & 10.851(030) & TNT \\
5834.45 & 19.97 & \nd & 11.814(024) & 10.866(025) & 10.830(019) & 10.866(027) & TNT \\
5835.45 & 20.97 & 11.527(573) & 11.918(008) & 10.952(046) & 10.786(042) & 10.852(066) & TNT \\
5836.45 & 21.97 & 12.200(044) & 12.001(013) & 11.011(023) & 10.859(017) & 10.827(043) & TNT \\
5837.45 & 22.97 & 12.126(065) & 12.155(009) & 11.023(037) & 10.851(013) & 10.787(019) & TNT \\
5839.45 & 24.97 & 12.663(068) & 12.243(009) & 11.196(027) & 10.920(017) & 10.743(035) & TNT \\
5840.45 & 25.97 & 12.637(042) & 12.310(010) & 11.234(021) & 10.949(013) & 10.749(024) & TNT \\
5841.48 & 27.00 & \nd & 12.422(011) & 11.305(050) & 11.032(035) & 10.747(025) & TNT \\
5847.44 & 32.96 & \nd & 12.888(016) & 11.679(054) & 11.348(050) & 11.056(095) & TNT \\
5848.45 & 33.97 & 12.938(058) & 12.942(013) & 11.749(010) & 11.457(024) & 11.089(023) & TNT \\
5850.44 & 35.96 & 13.152(065) & 13.045(013) & 11.854(022) & 11.598(022) & 11.237(025) & TNT \\
5851.44 & 36.96 & 13.091(060) & 13.048(026) & 11.930(012) & 11.646(020) & 11.310(023) & TNT \\
5852.45 & 37.97 & 13.309(042) & 13.106(017) & 11.948(013) & 11.677(015) & 11.365(028) & TNT \\
5858.43 & 43.95 & 13.358(064) & 13.281(009) & 12.152(019) & 11.936(013) & 11.686(022) & TNT \\
5891.90 & 77.42 & \nd & 13.747(009) & 13.078(010) & 13.031(010) & 13.119(021) & TNT \\
5894.87 & 80.39 & 14.305(050) & 13.791(008) & 13.157(017) & 13.114(014) & 13.218(021) & TNT \\
5906.89 & 92.41 & 14.728(044) & 13.941(011) & 13.451(010) & 13.497(011) & 13.645(020) & TNT \\
5908.91 & 94.43 & 14.700(048) & 13.956(011) & 13.466(015) & 13.555(012) & 13.707(020) & TNT \\
5910.93 & 96.45 & 14.837(053) & 13.962(009) & 13.547(008) & 13.621(015) & 13.777(022) & TNT \\
5911.86 & 97.38 & 14.958(055) & 14.007(011) & 13.592(011) & 13.653(011) & 13.815(021) & TNT \\
5912.93 & 98.45 & 15.056(043) & 14.035(011) & 13.601(011) & 13.684(012) &   13.862(019) &   TNT \\
5913.86 & 99.38 & 15.217(105) & 14.030(011) & 13.631(019) & 13.767(044) & 13.848(078) & TNT \\
5914.91 & 100.43 & 14.954(049) & 14.036(011) & 13.647(008) & 13.751(011) & 13.920(023) & TNT    \\
5916.91 & 102.43 & 15.230(045) & 14.076(011) & 13.689(008) & 13.803(011) & 13.973(018) & TNT    \\
5917.90 & 103.42 & 15.117(042) & 14.096(009) & 13.712(008) & 13.847(009) & 13.989(020) & TNT    \\
5918.91 & 104.43 & 15.219(042) & 14.101(009) & 13.728(006) & 13.862(010) & 14.036(019) & TNT    \\
5926.92 & 112.44 & 15.399(043) & 14.226(012) & 13.916(009) & \nd & 14.266(021) & TNT    \\
5928.88 & 114.40 & 15.593(046) & 14.268(010) & 13.955(009) & 14.122(014) & 14.296(021) & TNT    \\
5929.84 & 115.36 & 15.557(046) & 14.278(012) & 13.975(010) & 14.201(009) & 14.314(022) & TNT    \\
5934.85 & 120.37 & 15.768(048) & 14.359(015) & 14.108(010) & 14.390(011) & 14.522(018) & TNT    \\
5936.91 & 122.43 & 15.940(050) & 14.423(015) & 14.156(010) & 14.418(011) & 14.515(021) & TNT    \\
5937.92 & 123.44 & 15.883(048) & 14.426(012) & 14.195(009) & 14.496(011) & 14.598(019) & TNT    \\
5938.94 & 124.46 & \nd & 14.419(016) & 14.200(021) & 14.479(016) & 14.410(047) & TNT    \\
5954.92 & 140.44 & 16.606(051) & 14.673(012) & 14.481(012) & 14.849(017) & 14.833(027) & TNT    \\
5960.73 & 146.25 & 16.828(048) & 14.736(014) & 14.597(013) & 15.075(016) & 14.980(027) & TNT    \\
5965.84 & 151.36 & 16.537(059) & 14.857(011) & 14.711(013) & 15.208(013) & 15.175(023) & TNT    \\
5966.93 & 152.45 & \nd & 14.885(022) & 14.795(021) & 15.142(028) & 15.119(037) & TNT    \\
5967.90 & 153.42 & 16.728(085) & 14.860(013) & 14.781(013) & 15.234(012) & 15.155(022) & TNT    \\
5968.86 & 154.38 & 16.700(065) & 14.857(013) & 14.782(013) & 15.276(012) & 15.187(025) & TNT    \\
5971.89 & 157.41 & 17.024(056) & 14.911(013) & 14.856(014) & 15.284(017) & 15.252(023) & TNT    \\
5972.89 & 158.41 & 17.024(048) & 14.918(010) & 14.821(012) & 15.297(019) & 15.197(027) & TNT    \\
5979.85 & 165.37 & 17.228(050) & 15.056(013) & 14.900(011) & 15.511(024) & 15.282(029) & TNT    \\
5981.87 & 167.39 & 17.503(050) & 15.067(015) & 15.009(013) & 15.516(018) & 15.297(031) & TNT    \\
5982.84 & 168.36 & 17.248(061) & 15.073(017) & 14.990(017) & 15.574(023) & 15.371(037) & TNT    \\
5983.83 & 169.35 & 17.328(057) & 15.062(016) & 15.037(019) & 15.596(015) & 15.356(030) & TNT    \\
5985.88 & 171.40 & 17.274(077) & 15.125(013) & 15.112(011) & 15.699(014) & 15.520(024) & TNT    \\
5992.84 & 178.36 & 17.729(078) & 15.229(020) & 15.144(013) & 15.761(023) & 15.588(027) & TNT    \\
5994.86 & 180.38 & 17.248(097) & 15.421(041) & 15.176(036) & 16.037(076) & 15.558(081) & TNT    \\
5996.87 & 182.39 & 18.076(099) & 15.287(039) & 15.182(031) & 15.810(051) & 15.619(056) & TNT    \\
5998.85 & 184.37 & 17.510(081) & 15.327(014) & 15.282(011) & 15.950(014) & 15.672(022) & TNT    \\
5999.87 & 185.39 & 17.681(055) & 15.322(018) & 15.247(016) & 15.859(021) & 15.722(041) & TNT    \\
6010.82 & 196.34 & 18.074(062) & 15.477(014) & 15.500(013) & 16.201(017) & 15.798(025) & TNT    \\
6011.79 & 197.31 & \nd & 15.478(011) & 15.523(013) & 16.215(018) & 15.943(032) & TNT    \\
6012.72 & 198.24 & \nd & 15.474(018) & 15.505(014) & 16.092(033) & 15.789(056) & TNT    \\
6015.77 & 201.29 & 18.008(055) & 15.531(012) & 15.539(013) & 16.223(018) & 15.821(028) & TNT    \\
6020.80 & 206.32 & 17.974(081) & 15.676(012) & 15.666(015) & 16.355(025) & 16.057(029) & TNT    \\
6021.71 & 207.23 & \nd & 15.653(020) & 15.672(019) & 16.490(028) & 15.909(028) & TNT    \\
6029.75 & 215.27 & 18.140(065) & 15.784(013) & 15.785(012) & 16.511(015) & 16.153(021) & TNT    \\
6042.81 & 228.33 & 18.581(081) & 15.930(012) & 15.993(013) & 16.762(016) & 16.249(023) & TNT    \\
6050.62 & 236.14 & 18.052(278) & 16.094(033) & 16.102(035) & 16.909(066) & 16.363(051) & TNT    \\
6062.55 & 248.07 & 19.014(159) & 16.242(018) & 16.281(018) & 17.105(027) & 16.505(032) & TNT    \\
6069.62 & 255.14 & \nd & 16.304(038) & 16.374(019) & 16.968(077) & \nd & TNT    \\
6071.61 & 257.13 & \nd & 16.357(013) & 16.419(013) & 17.247(028) & 16.529(038) & TNT    \\
6077.52 & 263.04 & \nd & 16.588(114) & 16.443(043) & 17.325(065) & 16.855(087) & TNT    \\
6090.72 & 276.24 & \nd & 16.588(034) & 16.673(033) & 17.399(077) & 16.709(088) & TNT    \\
6108.53 & 294.05 & \nd & 17.318(139) & 17.234(082) & 17.850(117) & 16.937(061) & TNT    \\
6137.54 & 323.06 & \nd & 17.651(101) & 17.430(071) & 18.258(090) & 17.145(059) & TNT    \\
6160.59 & 346.11 & \nd & 17.918(140) & 17.814(107) & 18.051(155) & 18.096(260) & TNT    \\
6277.56 & 463.08 & \nd & \nd & 19.360(040) & 20.100(060) & 18.930(030) & LJT \\
\enddata
\tablenotetext{a}{2,450,000.5 has been subtracted from the Julian Date.}
\tablenotetext{b}{Relative to the epoch of $B$-band maximum (JD = 2,455,814.98).}
\tablenotetext{c}{Note: Uncertainties, in units of 0.001 mag, are $1\sigma$.}
\end{deluxetable}

\clearpage
\begin{deluxetable}{cccccc}
\tablecolumns{6} \tablewidth{0pc} \tabletypesize{\scriptsize}
\tablecaption{Journal of Spectroscopic Observations of SN 2011fe}
\tablehead{\colhead{UT Date} & \colhead{JD\tablenotemark{a}} &
\colhead{Phase\tablenotemark{b}} & \colhead{Exp.(s)} &
\colhead{Telescope + Instrument} & \colhead{Range (\AA)}} \startdata
2011 Aug. 25   &5798.55&$-$16   &2$\times$1800 & YNAO 2.4~m+YFOSC (G8+G14) &3500-9000 \\
2011 Aug. 26   &5809.54&$-$15   &2$\times$1800 & YNAO 2.4~m+YFOSC (G8+G14) &3500-9000 \\
2011 Aug. 27   &5800.53&$-$14   &2$\times$1200 & YNAO 2.4~m+YFOSC (G8+G14) &3500-9000 \\
2011 Aug. 28   &5801.53&$-$13   &2$\times$1200 & YNAO 2.4~m+YFOSC (G8+G14) &3500-9000 \\
2011 Aug. 28   &5801.54&$-$13   &1800          & BAO  2.16~m+BFOSC (G4)         &3400-8500 \\
2011 Aug. 29   &5802.53&$-$12   &2$\times$900  & YNAO 2.4~m+YFOSC (G8+G14) &3500-9000 \\
2011 Aug. 30    &5803.53&$-$11  &2$\times$600 & YNAO 2.4~m+YFOSC (G8+G14) &3500-9000 \\
2011 Aug. 31    &5804.54& $-$10  &2$\times$600 & YNAO 2.4~m+YFOSC (G8+G14) &3500-9000 \\
2011 Sept. 1    &5805.53& $-$9  &2$\times$600 & YNAO 2.4~m+YFOSC (G8+G14) &3500-9000 \\
2011 Sept. 2    &5806.50& $-$8  &1200         & BAO  2.16~m+OMR           &3400-8750 \\
2011 Sept. 2    &5806.52& $-$8  &2$\times$300 & YNAO 2.4~m+YFOSC (G8+G14) &3500-9000 \\
2011 Sept. 3    &5807.55& $-$7  &2$\times$300 & YNAO 2.4~m+YFOSC (G8+G14) &3500-9000  \\
2011 Sept. 4    &5808.55& $-$6  &600      & BAO  2.16~m+BFOSC (G4)         &4000-6700 \\
2011 Sept. 5    &5809.53& $-$5  &2$\times$300 & YNAO 2.4~m+YFOSC (G8+G14) &3500-9000 \\
2011 Sept. 5    &5809.54& $-$5  &600      & BAO  2.16~m+BFOSC (G4)         &4000-6700 \\
2011 Sept. 13   &5817.52& +3    &2$\times$480  & YNAO 2.4~m+YFOSC          &3500-9000 \\
2011 Sept. 17   &5821.48& +7    &300       & BAO  2.16~m+OMR           &4100-6900 \\
2011 Sept. 19   &5823.48& +9    &600       & BAO  2.16~m+OMR           &5500-6850 \\
2011 Sept. 19   &5823.52& +9    &2$\times$300 & YNAO 2.4~m+YFOSC (G8+G14) &3500-9000 \\
2011 Sept. 28   &5832.50& +18   &2$\times$300 & YNAO 2.4~m+YFOSC (G8+G14) &3500-9000 \\
2011 Sept. 30   &5834.45& +20   &1200      & BAO  2.16~m+OMR           &4500-8900 \\
2011 Oct. 3     &5837.45& +23   &600       & BAO  2.16~m+BFOSC (G4)        &3400-8700 \\
2011 Oct. 6     &5840.45& +26   &300       & BAO  2.16~m+BFOSC (G4)        &4000-6700 \\
2011 Dec. 3     &5898.93& +85   &900       & YNAO 2.4~m+YFOSC (G3)     & 3500-9000     \\
2011 Dec. 16    &5911.88& +97   &1800      & BAO  2.16~m+BFOSC(G4)     & 3300-9700     \\
2011 Dec. 23    &5918.92& +105  &1200      & BAO  2.16~m+OMR           &3800-9000\\
2011 Dec. 31    &5926.92& +113  &1200      & YNAO 2.4~m+YFOSC (G3)     &3350-9100\\
2012 Jan. 26    &5952.82& +138  &1800      & YNAO 2.4~m+YFOSC (G3)     &3900-9100 \\
2012 Feb. 05    &5962.89& +148  &2700      & YNAO 2.4~m+YFOSC (G3)     &3700-8300 \\
2012 Feb. 24    &5981.85& +167  &3600      & BAO  2.16~m+OMR           &3850-8500\\
2012 Mar. 16    &6002.71& +188  &3000      & BAO  2.16~m+OMR           &3500-8750\\
2012 Apr. 13    &6030.76& +216  &3600      & BAO  2.16~m+OMR           &4550-8200 \\
2012 Apr. 30    &6047.79& +233  &3000      & YNAO 2.4~m+YFOSC (G3)     &3400-9050 \\
2012 May. 23    &6070.79& +256  &3000      & YNAO 2.4~m+YFOSC (G3)     &3400-9050 \\
2012 Dec. 16    &6277.56& +463  &3600      & YNAO 2.4~m+YFOSC (G3)     &3400-9050 \\
\enddata
\tablenotetext{a}{2,450,000.5 has been subtracted from the Julian Date.}
\tablenotetext{b}{Relative to the epoch of $B$-band maximum (JD = 2,455,814.98).}
\end{deluxetable}

\begin{deluxetable}{rl}
\tablecolumns{6} \tablewidth{0pc} \tabletypesize{\scriptsize}
\tablecaption{Photometric Parameters of SN 2011fe}
\tablehead{\colhead{Parameters} & \colhead{Value}} \startdata
$t_{B_{\rm max}}$ (JD) & 2,455,814.48 $\pm$ 0.03 \\
$B_{\rm max}$ (mag) & $9.96 \pm 0.03$ \\
$\Delta m_{15}$ (mag) & $1.18 \pm 0.03$ \\
$B_{\rm max}-V_{\rm max}$ (mag) & $-0.03 \pm 0.04$  \\
Host galaxy & M101 \\
Absolute magnitude $U$ &  $-19.74 \pm 0.19$ \\
$B$ & $-19.23 \pm 0.19$ \\
$V$& $-19.17 \pm 0.19$ \\
$R$ & $-19.11 \pm 0.19$\\
$I$ & $-18.85 \pm 0.19$ \\
$E(B-V)_{\rm MW}$ (mag) & 0.008 \\
$E(B-V)_{\rm host}$ (mag) & $0.032 \pm 0.045$ \\
Late-time decline rate (mag/100 days) $U$ & $2.28 \pm 0.06$ \\
$B$ & $1.44 \pm 0.02$ \\
$V$ &  $1.51 \pm 0.02$ \\
$R$ & $1.71 \pm 0.03$ \\
$I$ & $1.12 \pm 0.04$ \\
\enddata
\end{deluxetable}

\begin{deluxetable}{ccccccc}
\tablecolumns{5} \tablewidth{0pc} \tabletypesize{\scriptsize}
\tablecaption{Bolometric Light Curve of SN 2011fe}
\tablehead{Phase\tablenotemark{a} & \colhead{log($L$) [erg
s$^{-1}$]\tablenotemark{b}} & Phase & \colhead{log($L$) [erg
s$^{-1}$]} } \startdata
-16.31 & 41.09 &  20 & 42.51\\
-16 & 41.26 &  21 & 42.49\\
-15 & 41.63  & 22 & 42.45\\
-14 & 41.90 &  23 & 42.45\\
-13 & 42.09  & 25 & 42.41\\
-12 & 42.31 &  26 & 42.40\\
-11 & 42.49 &  27 & 42.38\\
-10 & 42.62 &  33 & 42.25\\
-9 & 42.71  & 37 & 42.15\\
-8 & 42.79  & 44 & 42.03\\
-7 & 42.89  & 49 & 41.96\\
-6 & 42.92 &  55 & 41.88\\
-5 & 42.99  & 60 & 41.82\\
-4 & 43.00  & 64 & 41.77\\
-3 & 43.03  & 80. & 41.57\\
-2 & 43.05 &  92. & 41.45\\
-1 & 43.05 &  97. & 41.40\\
1 & 43.05  & 104 & 41.33\\
2 & 43.04  & 114 & 41.25\\
3 & 43.00  & 120 & 41.18\\
4 & 43.00  & 123 & 41.15\\
5 & 43.00 &  140 & 41.03\\
7 & 42.95 &  151 & 40.94\\
8 & 42.91 &  158 & 40.91\\
9 & 42.85 &  165 & 40.86\\
10 & 42.78 &  171 & 40.80\\
11 & 42.74 &  178 & 40.77\\
12 & 42.72  & 185 & 40.73\\
13 & 42.68  & 196 & 40.65\\
14 & 42.65 &  206 & 40.58\\
15 & 42.61 &  215 & 40.53\\
16 & 42.58  & 228 & 40.46\\
18 & 42.58 &  236 & 40.41\\
19 & 42.53 &  248 & 40.34\\
\enddata
\tablenotetext{a}{Relative to the epoch of $B$-band maximum (JD = 2,455,814.98).}
\tablenotetext{b}{Typical uncertainty is 0.07, dominated by the uncertainty in the distance.}
\end{deluxetable}

\begin{deluxetable}{ccccccc}
\tablecolumns{5} \tablewidth{0pc} \tabletypesize{\scriptsize}
\tablecaption{Fit Parameters of the Rising Light Curve}
\tablehead{\colhead{Filter} & \colhead{Data points}  &
\colhead{$t_{\rm first}$($f\propto t^2$)
[days]\tablenotemark{a}\tablenotemark{b}} &
\colhead{reduced~$\chi^2$} & \colhead{$t_{\rm first}$($f\propto t^n$)
[days]} & \colhead{$n$} & \colhead{reduced~$\chi^2$}} \startdata
$U$ & 12 & $-17.37(01)$ & 31.61 & $-19.37(21)$ & $3.42(14)$ & 4.35\\%
$B$ & 13 & $-17.46(01)$ & 19.18 & $-18.50(11)$ & $2.76(08)$ & 4.76 \\%
$V$ & 13 & $-17.82(02)$ & 10.41 & $-18.24(10)$ & $2.27(06)$ & 9.26\\%
$R$ & 13 & $-17.76(02)$ & 6.98 & $-18.36(11)$ & $2.39(07)$ & 3.39\\%
$I$  & 13 & $-17.66(02)$ & 3.49 & $-18.12(14)$ & $2.33(09)$ & 2.26\\ \\%
\hline
& & & Multi-band Fitting  & $UBVRI$ bands& & \\ \\%
\hline
$U$ & 12 & $-17.59(01)$ & 52.52  & $-18.18(29)$ & $2.63(16)$ & 10.65\\%
$B$ & 12 & $-17.59(01)$ & 28.52 & $-18.18(29)$  & $2.56(12)$ & 1.40 \\%
$V$ & 12 & $-17.59(01)$ & 18.35 & $-18.18(29)$ & $2.25(21)$ & 1.50 \\%
$R$ & 12 & $-17.59(01)$ & 15.01 & $-18.18(29)$ & $2.29(34)$ & 1.62 \\%
$I$ & 12 & $-17.59(01)$ & 3.91 & $-18.18(29)$ & $2.37(42)$ & 2.29 \\
Total reduced $\chi^2$ & 60 &  $-17.59(01)$   & 21.91 &  $-18.18(29)$ & \nd  & 3.21\\\\%
\hline
& & & Multi-band Fitting  & $BVRI$ bands & & \\ \\%
\hline
$B$ & 12 & $-17.63(01)$ & 38.75 & $-18.00(16)$ & $2.43(09)$ & 2.23 \\%
$V$ & 12 & $-17.63(01)$ & 10.33 & $-18.00(16)$ & $2.14(07)$ & 1.01\\%
$R$ & 12 & $-17.63(01)$ & 9.12 & $-18.00(16)$ & $2.17(11)$ & 2.25 \\%
$I$  & 12 & $-17.63(01)$ & 3.30 & $-18.00(16)$& $2.25(09)$ & 2.24\\ %
Total reduced $\chi^2$ & 48 & $-17.63(01)$ & 14.30 &  $-18.00(16)$ & \nd  & 1.78 \\\\%
\enddata
\tablenotetext{a}{Relative to the epoch of $B$-band maximum (JD = 2,455,814.98).}
\tablenotetext{b}{Note: Uncertainties, in units of 0.01, are $1\sigma$.}
\tablenotetext{c}{$t^2$ model cannot fit these bolometric data reasonably.}
\end{deluxetable}


\begin{thebibliography}{}

\bibitem[Arnett(1982)]{1982ApJ...253..785A} Arnett, W.~D.\ 1982, \apj, 253, 785
\bibitem[Baron et al.(2015)]{2015MNRAS.454.2549B} Baron, E., H\"{o}flich, P., Friesen, B., et al.\ 2015, \mnras, 454, 2549
\bibitem[Benetti et al.(2005)]{2005ApJ...623.1011B} Benetti, S., Cappellaro, E., Mazzali, P.~A., et al.\ 2005, \apj, 623, 1011
\bibitem[Bloom et al.(2012)]{2012ApJ...744L..17B} Bloom, J.~S., Kasen, D., Shen, K.~J., et al.\ 2012, \apjl, 744, L17
\bibitem[Bresolin(2007)]{2007ApJ...656..186B} Bresolin, F.\ 2007, \apj, 656, 186
\bibitem[Brown et al.(2015)]{2015ApJ...809...37B} Brown, P.~J., Baron, E., Milne, P., Roming, P.~W.~A., \& Wang, L.\ 2015, \apj, 809, 37
\bibitem[Brown et al.(2012)]{2012ApJ...753...22B} Brown, P.~J., Dawson, K.~S., de Pasquale, M., et al.\ 2012, \apj, 753, 22
\bibitem[Chomiuk et al.(2012)]{2012ApJ...750..164C} Chomiuk, L., Soderberg, A.~M., Moe, M., et al.\ 2012, \apj, 750, 164
\bibitem[Conley et al.(2006)]{2006AJ....132.1707C} Conley, A., Howell, D.~A., Howes, A., et al.\ 2006, \aj, 132, 1707
\bibitem[Cousins(1981)]{1981SAAOC...6....4C} Cousins, A.~W.~J.\ 1981, South African Astronomical Observatory Circular, 6, 4
\bibitem[Fan et al.(2015)]{2015RAA....15..918F} Fan, Y.-F., Bai, J.-M., Zhang, J.-J., et al.\ 2015, Research in Astronomy and Astrophysics, 15, 918
\bibitem[Filippenko et al.(2001)]{2001ASPC..246..121F} Filippenko, A.~V., Li, W.~D., Treffers, R.~R., \& Modjaz, M.\ 2001, in Small-Telescope Astronomy on Global Scales, ed. W. P. Chen, C. Lemme, \& B. Paczy\'{n}ski (San Francisco: ASP Vol. 246), 121
\bibitem[Firth et al.(2015)]{2015MNRAS.446.3895F} Firth, R.~E., Sullivan, M., Gal-Yam, A., et al.\ 2015, \mnras, 446, 3895
\bibitem[Foley et al.(2012)]{2012ApJ...744...38F} Foley, R.~J., Challis, P.~J., Filippenko, A.~V., et al.\ 2012, \apj, 744, 38
\bibitem[Foley \& Kirshner(2013)]{2013ApJ...769L...1F} Foley, R.~J., \& Kirshner, R.~P.\ 2013, \apjl, 769, L1
\bibitem[Fransson \& Jerkstrand(2015)]{2015arXiv151100245F} Fransson, C., \& Jerkstrand, A.\ 2015, arXiv:1511.00245
\bibitem[Ganeshalingam et al.(2011)]{2011MNRAS.416.2607G} Ganeshalingam, M., Li, W., \& Filippenko, A.~V.\ 2011, \mnras, 416, 2607
\bibitem[Graham et al.(2015b)]{2015MNRAS.446.2073G} Graham, M.~L., Foley, R.~J., Zheng, W., et al.\ 2015b, \mnras, 446, 2073
\bibitem[Graham et al.(2015a)]{2015MNRAS.454.1948G} Graham, M.~L., Nugent, P.~E., Sullivan, M., et al.\ 2015a, \mnras, 454, 1948
\bibitem[H{\"o}flich et al.(1998)]{1998ApJ...495..617H} H{\"o}flich, P.,  Wheeler, J.~C., \& Thielemann, F.~K.\ 1998, \apj, 495, 617
\bibitem[Horesh et al.(2012)]{2012ApJ...746...21H} Horesh, A., Kulkarni, S.~R., Fox, D.~B., et al.\ 2012, \apj, 746, 21
\bibitem[Huang et al.(2012)]{2012RAA....12.1585H} Huang, F., Li, J.-Z., Wang, X.-F., et al.\ 2012, Research in Astronomy and Astrophysics, 12, 1585
\bibitem[Iben \& Tutukov(1984)]{1984ApJS...54..335I} Iben, I., Jr., \& Tutukov, A.~V.\ 1984, \apjs, 54, 335
\bibitem[Johnson et al.(1966)]{1966CoLPL...4...99J} Johnson, H.~L., Mitchell, R.~I., Iriarte, B., \& Wisniewski, W.~Z.\ 1966, Communications of the Lunar and Planetary Laboratory, 4, 99
\bibitem[Kasen(2010)]{2010ApJ...708.1025K} Kasen, D.\ 2010, \apj, 708, 1025
\bibitem[Leloudas et al.(2009)]{2009A&A...505..265L} Leloudas, G., Stritzinger, M.~D., Sollerman, J., et al.\ 2009, \aap, 505, 265
\bibitem[Lentz et al.(2000)]{2000ApJ...530..966L} Lentz, E.~J., Baron, E., Branch, D., Hauschildt, P.~H., \& Nugent, P.~E.\ 2000, \apj, 530, 966
\bibitem[Li et al.(2011)]{2011Natur.480..348L} Li, W., Bloom, J.~S., Podsiadlowski, P., et al.\ 2011, \nat, 480, 348
\bibitem[Li et al.(2002)]{2002PASP..114..403L} Li, W., Filippenko, A.~V., Van Dyk, S.~D., et al.\ 2002, \pasp, 114, 403
\bibitem[Lundqvist et al.(2015)]{2015A&A...577A..39L} Lundqvist, P., Nyholm, A., Taddia, F., et al.\ 2015,\aap, 577, A39
\bibitem[Marion et al.(2015)]{2015arXiv150707261M} Marion, G.~H., Brown, P. ~J., Vink{\'o}, J., et al.\ 2015, arXiv:1507.07261
\bibitem[Matheson et al.(2012)]{2012ApJ...754...19M} Matheson, T., Joyce, R.~R., Allen, L.~E., et al.\ 2012, \apj, 754, 19
\bibitem[Mazzali et al.(2015)]{2015MNRAS.450.2631M} Mazzali, P.~A., Sullivan, M., Filippenko, A.~V., et al.\ 2015, \mnras, 450, 2631
\bibitem[McClelland et al.(2013)]{2013ApJ...767..119M} McClelland, C.~M., Garnavich, P.~M., Milne, P.~A., Shappee, B.~J., \& Pogge, R.~W.\ 2013, \apj, 767, 119
\bibitem[Miles et al.(2015)]{2015arXiv150805961M} Miles, B.~J., van Rossum, D.~R., Townsley, D.~M., et al.\ 2015, arXiv:1508.05961
\bibitem[Milne et al.(1999)]{1999ApJS..124..503M} Milne, P.~A., The, L.-S., \& Leising, M.~D.\ 1999, \apjs, 124, 503
\bibitem[Milne et al.(2001)]{2001ApJ...559.1019M} Milne, P.~A., The, L.-S., \& Leising, M.~D.\ 2001, \apj, 559, 1019
\bibitem[Munari et al.(2013)]{2013NewA...20...30M} Munari, U., Henden, A., Belligoli, R., et al.\ 2013, \na, 20, 30
\bibitem[Nugent et al.(2011)]{2011Natur.480..344N} Nugent, P.~E., Sullivan, M., Cenko, S.~B., et al.\ 2011, \nat, 480, 344
\bibitem[Parrent et al.(2012)]{2012ApJ...752L..26P} Parrent, J.~T., Howell, D.~A., Friesen, B., et al.\ 2012, \apjl, 752, L26
\bibitem[Patat et al.(2013)]{2013A&A...549A..62P} Patat, F., Cordiner, M.~A., Cox, N.~L.~J., et al.\ 2013, \aap, 549, A62
\bibitem[Pereira et al.(2013)]{2013A&A...554A..27P} Pereira, R., Thomas, R.~C., Aldering, G., et al.\ 2013, \aap, 554, A27
\bibitem[Phillips et al.(1999)]{1999AJ....118.1766P} Phillips, M.~M., Lira, P., Suntzeff, N.~B., et al.\ 1999, \aj, 118, 1766
\bibitem[Piro et al.(2010)]{2010ApJ...708..598P} Piro, A.~L., Chang, P., \& Weinberg, N.~N.\ 2010, \apj, 708, 598
\bibitem[Piro \& Nakar(2013)]{2013ApJ...769...67P} Piro, A.~L., \& Nakar, E.\ 2013, \apj, 769, 67
\bibitem[Piro \& Nakar(2014)]{2014ApJ...784...85P} Piro, A.~L., \& Nakar, E.\ 2014, \apj, 784, 85
\bibitem[Richmond \& Smith(2012)]{2012JAVSO..40..872R} Richmond, M.~W., \& Smith, H.~A.\ 2012, Journal of the American Association of Variable Star Observers (JAAVSO), 40, 872
\bibitem[Riess et al.(1999)]{1999AJ....118.2675R} Riess, A.~G., Filippenko, A.~V., Li, W., et al.\ 1999, \aj, 118, 2675
\bibitem[Sauer et al.(2008)]{2008MNRAS.391.1605S} Sauer, D.~N., Mazzali, P.~A., Blondin, S., et al.\ 2008, \mnras, 391, 1605
\bibitem[Schlafly \& Finkbeiner(2011)]{2011ApJ...737..103S} Schlafly, E.~F., \& Finkbeiner, D.~P.\ 2011, \apj, 737, 103
\bibitem[Shappee et al.(2015)]{2015arXiv150704257S} Shappee, B.~J., Piro, A.~L., Holoien, T.~W.-S., et al.\ 2015, arXiv:1507.04257
\bibitem[Shappee \& Stanek(2011)]{2011ApJ...733..124S} Shappee, B.~J., \& Stanek, K.~Z.\ 2011, \apj, 733, 124
\bibitem[Shappee et al.(2013)]{2013ApJ...762L...5S} Shappee, B.~J., Stanek, K.~Z., Pogge, R.~W., \& Garnavich, P.~M.\ 2013, \apjl, 762, L5
\bibitem[Silverman et al.(2015)]{2015MNRAS.451.1973S} Silverman, J.~M., Vink{\'o}, J., Marion, G.~H., et al.\ 2015, \mnras, 451, 1973 (S15)
\bibitem[Smith et al.(2011)]{2011arXiv1111.6626S} Smith, P.~S., Williams, G.~G., Smith, N., et al.\ 2011, arXiv:1111.6626
\bibitem[Sollerman et al.(1998)]{1998A&A...337..207S} Sollerman, J., Leibundgut, B., \& Spyromilio, J.\ 1998, \aap, 337, 207
\bibitem[Song et al.(2016)]{} Song, H., et al. 2016, in prep.
\bibitem[Stanishev et al.(2007)]{2007A&A...469..645S} Stanishev, V., Goobar, A., Benetti, S., et al.\ 2007, \aap, 469, 645
\bibitem[Stoll et al.(2011)]{2011ATel.3588....1S} Stoll, R., Shappee, B., \& Stanek, K.~Z.\ 2011, The Astronomer's Telegram, 3588, 1
\bibitem[Stritzinger \& Leibundgut(2005)]{2005A&A...431..423S} Stritzinger, M., \& Leibundgut, B.\ 2005, \aap, 431, 423
\bibitem[Tammann \& Reindl(2011)]{2011arXiv1112.0439T} Tammann, G.~A., \& Reindl, B.\ 2011, arXiv:1112.0439
\bibitem[Timmes et al.(2003)]{2003ApJ...590L..83T} Timmes, F. X., Brown, Edward F.,\& Truran, J. W.\ 2003, \apjl, 590, L83
\bibitem[Taubenberger et al.(2015)]{2015MNRAS.448L..48T} Taubenberger, S., Elias-Rosa, N., Kerzendorf, W.~E., et al.\ 2015, \mnras, 448, L48
\bibitem[Thomas et al.(2011)]{2011PASP..123..237T} Thomas, R.~C., Nugent, P.~E., \& Meza, J.~C.\ 2011, \pasp, 123, 237
\bibitem[Wang et al.(2009b)]{2009ApJ...699L.139W} Wang, X., Filippenko, A.~V., Ganeshalingam, M., et al.\ 2009b, \apjl, 699, L139
\bibitem[Wang et al.(2008)]{2008ApJ...675..626W} Wang, X., Li, W., Filippenko, A.~V., et al.\ 2008, \apj, 675, 626
\bibitem[Wang et al.(2009a)]{2009ApJ...697..380W} Wang, X., Li, W., Filippenko, A.~V., et al.\ 2009a, \apj, 697, 380
\bibitem[Wang et al.(2013)]{2013Sci...340..170W} Wang, X., Wang, L., Filippenko, A.~V., Zhang, T., and Zhao, X.\ 2013, Science, 340, 170
\bibitem[Whelan \& Iben(1973)]{1973ApJ...186.1007W} Whelan, J., \& Iben, I., Jr.\ 1973, \apj, 186, 1007
\bibitem[Zhang et al. (2010)]{2010PASP..122....1Z} Zhang, T., Wang, X., Li, W., et al. \ 2010, \pasp, 122, 1
\bibitem[Zhang et al.(2016)]{} Zhang, J., et al. 2016, in prep.
\bibitem[Zhao et al.(2015)]{2015ApJS..220...20Z} Zhao, X., Wang, X., Maeda, K., et al.\ 2015, \apjs, 220, 20 (Z15)
\bibitem[Zhao et al.(2016)]{} Zhao, X., et al. 2016, in prep.
\bibitem[Zheng et al.(2013)]{2013ApJ...778L..15Z} Zheng, W., Silverman, J.~M., Filippenko, A.~V., et al.\ 2013, \apjl, 778, L15

\end{thebibliography}
\end{document}